\documentclass[11pt,a4paper]{article}
\pdfoutput=1
\usepackage{jheppub}
\usepackage[utf8]{inputenc}
\usepackage{epstopdf}
\usepackage{tabularx}
\usepackage{amsmath,amssymb, amsfonts,amsbsy,mathrsfs,graphicx, booktabs, slashed, xcolor, adjustbox}
\usepackage{latexsym}
\usepackage{hyperref}
\usepackage{bm}
\usepackage{comment}

\usepackage{bbm}
\usepackage{scalerel}
\newcommand{\pmx}[1]{\begin{pmatrix} #1 \end{pmatrix}}
\newcommand{\zero}{_{\scriptscriptstyle 0}}
\newcommand{\zeroone}{_{\scriptscriptstyle 0,1}}
\newcommand{\zerotwo}{_{\scriptscriptstyle 0,2}}
\newcommand{\zzero}[1]{_{{\scriptscriptstyle 0}, #1}}
\newcommand{\transpose}{{\scaleto{\mathsf{T}}{5pt}}}
\newcommand{\bas}{_\phi}
\newcommand{\hd}{^\text{h.d.}}
\newcommand{\gen}{^{\prime}}
\newcommand{\dcamp}{
    \mathcal{M}\gen
}

\newcommand*{\citeword}{}

\def\equationautorefname~#1\null{Eq.\,(#1)\null}
\def\sectionautorefname~#1\null{Sec.\,#1\null}
\def\subsectionautorefname~#1\null{Sec.\,#1\null}
\def\figureautorefname~#1\null{Fig.\,#1\null}
\def\appendixautorefname~#1\null{App.\,#1\null}

\makeatletter\g@addto@macro\bfseries{\boldmath}\makeatother

\newcommand{\angl}[2]{\langle #1 #2 \rangle} 
\newcommand{\sqr}[2]{\left[ #1 #2 \right]}

\newcommand{\Tr}{\mathrm{Tr}}

\newcommand{\cref}[1]{Chapter~\ref{ch:#1}}

\newcommand{\nn}{\nonumber }

\newcommand{\beq}{\begin{equation}} 
\newcommand{\eeq}{\end{equation}} 
\newcommand{\ba}{\begin{array}}  
\newcommand{\ea}{\end{array}} 
\newcommand{\bea}{\begin{eqnarray}}  
\newcommand{\eea}{\end{eqnarray} }  
\newcommand{\be}{\begin{eqnarray}}  
\newcommand{\ee}{\end{eqnarray} }  
\newcommand{\bal}{\begin{align}}
\newcommand{\eal}{\end{align}}   
\newcommand{\ben}{\begin{enumerate}}  
\newcommand{\een}{\end{enumerate}}  
\newcommand{\bc}{\begin{center}}
\newcommand{\ec}{\end{center}} 
\newcommand{\bt}{\begin{table}}
\newcommand{\et}{\end{table}}  
\newcommand{\btb}{\begin{tabular}}
\newcommand{\etb}{\end{tabular}}

\renewcommand{\[}{\left[}
\renewcommand{\]}{\right]}
\renewcommand{\(}{\left(}
\renewcommand{\)}{\right)}
\def\bes{\begin{equation*}}
\def\ees{\end{equation*}}
\def\bead{\begin{aligned}}
\def\eead{\end{aligned}}
\def\bmat{\left(\begin{matrix}}
\def\emat{\end{matrix}\right)}

\newcommand{\cO}{{\mathcal O}}

\newcommand{\cA}{{\mathcal A}}   

\newcommand{\tr}{\mathrm T \mathrm r}

\newcommand{\eps}{\epsilon}

\author[a]{Quentin Bonnefoy,}
\author[b]{Gauthier Durieux,}
\author[a,c]{Christophe Grojean,}
\author[a]{Camila S. Machado,}
\author[c]{Jasper Roosmale Nepveu}

\affiliation[a]{Deutsches Elektronen-Synchrotron DESY, Notkestr. 85, 22607 Hamburg, Germany}
\affiliation[b]{CERN, Theoretical Physics Department, Geneva 23 CH-1211, Switzerland}
\affiliation[c]{Humboldt-Universit\"at zu Berlin, Institut f\"ur Physik, Newtonstr. 15, 12489 Berlin, Germany}
\emailAdd{quentin.bonnefoy@desy.de}
\emailAdd{gauthier.durieux@cern.ch}
\emailAdd{christophe.grojean@desy.de}
\emailAdd{camila.machado@desy.de}
\emailAdd{jasper.roosmalenepveu@physik.hu-berlin.de}

\title{
The seeds of EFT double copy
}
\abstract{
We explore the double copy of effective field theories (EFTs), in the recently proposed generalized color-kinematics and Kawai--Lewellen--Tye (KLT) approaches.
In the former, we systematically construct scalar numerators satisfying the Jacobi identities from simpler numerator \emph{seeds} with trace-like permutation properties.
This construction has the advantage of being easily applicable to any multiplicity, which we exemplify up to 6-point.
It employs the linear map between color factors formed by single traces of generators and by products of the structure constants, which also relates the generalized KLT and color-kinematics formalisms, allowing to produce KLT kernels at arbitrary order in the EFT expansion.
At 4-point, we show that all EFT kernels are generated and that they only yield double-copy amplitudes which can also be obtained from the traditional KLT kernel.
We perform initial checks suggesting that the same conclusions also hold at 5-point. We focus on single-trace massless scalar EFTs which however also control the higher-derivative corrections to gauge and gravity theories.
}

\begin{document}
\begin{flushright}
DESY-21-223\\
CERN-TH-2021-224\\
HU-EP-21/59-RTG
\end{flushright}
\flushbottom

\makeatletter\renewcommand{\@fpheader}{\ }\makeatother
\maketitle

\section{Introduction}

Gravity amplitudes can be obtained from Yang--Mills amplitudes through a squaring operation  known as the \emph{double copy}.
This relation between theories has been studied through different lenses, exhibiting a rich underlying mathematical structure.
New elements of understanding are however still being collected.
The double copy was first discovered by Kawai, Lewellen and Tye (KLT) from relations between tree-level open- and closed-string amplitudes \cite{Kawai:1985xq}.
In the KLT formalism, (dilaton-axion-)gravity amplitudes are obtained from products of the appropriate color-ordered on-shell amplitudes of a gauge theory, normalized by a  scalar factor dubbed the KLT \emph{kernel}, which depends on the kinematics of the process.

A different but equivalent product structure was discovered at the level of trivalent graphs in field theory by Bern, Carrasco and Johansson (BCJ) \cite{Bern:2008qj,Bern:2010ue}.
In this approach, each Yang--Mills graph is expressed in terms of a color and a kinematic \emph{numerator} multiplying propagators.
It was found that both types of numerators can be chosen such that they obey identical algebraic relations under particle permutations, e.g.\ Jacobi-like identities.
This remarkable fact is known as the color-kinematics (CK) duality.
Replacing the color numerator by a second kinematic numerator then yields gravity amplitudes, in an economical way.
The double-copy structure was also observed using scattering equations by Cachazo, He and Yuan~\cite{Cachazo:2013gna, Cachazo:2013hca, Cachazo:2013iea}, who identified the aforementioned KLT kernel with the inverse matrix of color-ordered amplitudes computed in a bi-adjoint scalar theory.

In recent years, it has become clear that double-copy relations exist beyond the ({gravity}) = ({Yang--Mills})$^2$ example, and that theories different from Yang--Mills can be used as inputs, thereby double-copying to other theories than gravity.
A web of theories connected through double-copy relations was identified and further explorations are ongoing about the space of theories it covers \cite{Cachazo:2013iea,Carrasco:2019qwr}.
It was for example found that gauge theories with massive (scalar or fermionic) matter in the fundamental representation can also be double-copied 
\cite{Johansson:2014zca,
Johansson:2015oia,
delaCruz:2016wbr,
Brown:2016hck,
Brown:2018wss,
Johansson:2019dnu,
Plefka:2019wyg}, and even theories with spontaneous symmetry breaking obey the CK duality \cite{Chiodaroli:2015rdg}.
In addition, the double copy of massive gauge bosons has been considered in
\cite{Johnson:2020pny,
Momeni:2020vvr,
Momeni:2020hmc,
Hang:2021fmp,
Li:2021yfk}.
Besides gauge theories, a double-copy structure has been observed in pure scalar theories such as the aforementioned bi-adjoint scalar theory and the non-linear sigma model~\cite{Chen:2013fya,Cachazo:2014xea}.
Furthermore, the double-copy product based on the CK duality has been found to extend to the loop level in various examples.
For a comprehensive review on the double copy, see \citeword\cite{Bern:2019prr}.

The above examples mostly correspond to renormalizable theories, however higher-dimensional operators are also expected to take part in some form of the double copy.
Higher-derivative corrections to Yang--Mills theories were explicitly studied in \citeword\cite{Broedel:2012rc} and more recently in \citeword\cite{Menezes:2021dyp}, while scalar effective field theories (EFTs) were studied in e.g.\
\cite{Elvang:2018dco, 
Elvang:2020kuj, 
CarrilloGonzalez:2019fzc,
Kampf:2021jvf,
Low:2019wuv,Low:2020ubn}.
Other heavy mass EFTs have also been considered in \cite{Brandhuber:2021kpo,Haddad:2020tvs}.

Two more systematic approaches to study the range of higher-derivative operators that can be double-copied have recently been proposed by Carrasco, Rodina, Yin, Zekio\u{g}lu~\cite{Carrasco:2019yyn,Carrasco:2021ptp} and Chi, Elvang, Herderschee, Jones, Paranjape \cite{Chi:2021mio}.
In the former, the color-kinematics approach to the double copy is taken and extended by considering \emph{generalized numerators} that may simultaneously depend on both color and kinematics, while still satisfying Jacobi-like identities.
A set of composition rules is defined to systematically build such numerators at a given order in the EFT expansion.
This approach can be extended from 4- to 5-point amplitudes \cite{Carrasco:2021ptp}.
 
In the approach taken by \cite{Chi:2021mio}, the KLT kernel is generalized.
The initial observation is that the map between open- and closed-string amplitudes involves a kernel with higher-derivative corrections, which is the inverse matrix of the color-ordered  amplitudes in a bi-adjoint EFT with operator coefficients correlated in a specific way \cite{Mizera:2016jhj}.
This notion is generalized by allowing for more free parameters in the KLT kernel, arising from a more general bi-adjoint EFT. 
A set of bootstrap equations on the KLT kernel and input theories are proposed to guarantee a healthy analytical structure in the resulting double-copy amplitudes.
Solving the bootstrap then yields a systematic study of the operators possibly involved in the double copy. 
Correct factorization properties for a theory with fixed particle content need to be imposed as an additional constraint on both the kernel and the input amplitudes.

In this paper, we aim to shed light on these two approaches to the double copy of EFTs and on their relation, by introducing a new method to construct generalized numerators at any multiplicity. 
The traditional color numerators, consisting of products of Lie-group structure constants, can be written as linear combinations of the single traces of products of group generators.
In the same way, we show that all generalized numerators can be constructed from simple \textit{numerator seeds}, which satisfy the same permutation properties as the single traces of generators.

The construction of numerators has previously been investigated from different perspectives.
They have for instance been extracted from known amplitudes and the KLT kernel~\cite{Bjerrum-Bohr:2010pnr,KiermaierTalk2010}.
\emph{Dual trace factors}, analogous to our numerator seeds but involving momenta and polarization vectors, have also previously been identified in Yang--Mills amplitudes~\cite{Bern:2011ia, Bjerrum-Bohr:2012kaa, Du:2013sha, Fu:2013qna,Naculich:2014rta}.
In contrast, we use numerators to construct EFT amplitudes, and seeds built out of color and momenta to construct generalized scalar numerators.
Based on the kinematic algebra, vector numerators for Yang--Mills and heavy-quark effective theories have also recently been constructed from simpler ``pre-numerators''~\cite{Brandhuber:2021kpo, Brandhuber:2021bsf}.

A further advantage of numerator seeds is that they can be directly related to KLT kernels. This enables the study of the operators involved in the KLT double copy, through a method that is alternative to the bootstrap of \cite{Chi:2021mio}.
Numerator seeds thus provide further insight into the relation between the double copy approaches of \cite{Carrasco:2019yyn,Carrasco:2021ptp} and \cite{Chi:2021mio}, and into the structure of the generalized KLT kernel.
In particular, at 4-point, we show that the double-copy amplitudes obtained with a generalized KLT kernel can equivalently be achieved by the traditional kernel, multiplying healthy local input amplitudes including higher-derivative corrections. 
As emphasized in \cite{Chi:2021mio}, the generalized kernel does however allow for more general EFT inputs to the double copy.
We also report on various results which indicate that this observation extends to higher multiplicities.

The structure of this paper is as follows. 
To be self-contained and set the notation, we first provide in \autoref{sec:review} a detailed review of the two aforementioned approaches to the double copy of EFTs.
The construction of generalized numerators from seeds at any multiplicity is then presented in \autoref{sec:seeds}.
In this section, we also show how this construction facilitates the reorganization of CK-dual representations of amplitudes in terms of color-ordered amplitudes, which are the building blocks of the KLT formalism.
Moreover, for any input amplitude that can be double-copied with a generalized kernel, we identify new objects which yield the same double copy with the traditional kernel.
This holds provided the generalized kernel can be constructed from numerator seeds, and provided one can ensure locality of the new objects in order to call them amplitudes.
We discuss these two caveats at 4- and 5-point in the subsequent sections.
Restricting to 4-point amplitudes, \autoref{4ptsCKSection} and \autoref{s:doublecopystructure} illustrate our method and show that it generates all solutions to the KLT bootstrap.
We also analyse the double-copy structure in the KLT formalism, and the factorization properties of the amplitudes involved.
The two caveats above are successfully addressed in this 4-point case.
Moving on to 5-point amplitudes in \autoref{5pts}, we demonstrate that the lowest-order bootstrap solutions provided in~\cite{Chi:2021mio}, can be reproduced from our numerator construction.
We also present partial results suggesting that no new double copies are generated by the generalized kernel at 5-point either.

\section{The systematic double copy of EFTs}
\label{sec:review}

We start with a review on the generalized KLT method of \cite{Chi:2021mio} and the generalized numerator method of~\cite{Carrasco:2019yyn,Carrasco:2021ptp}.

\subsection{The generalized KLT approach} \label{KLTreview}
The KLT formula for an amplitude with $n$ external particles in the adjoint representation of $\text{SU}(N)$ (or $\text{U}(N)$) symmetry groups is given by
\begin{equation}
\label{eq:KLT}
\mathcal{M}_n
= \sum_{\alpha, \beta}^{(n-3)!}
	{A}_{n}^{L }[\alpha] \;
	S_n[\alpha | \beta] \;
	{A}_{n}^{ R}[\beta]\, . 
\end{equation}
Here, ${A}_{n}^{L}$ and $A_n^R$ are the color-ordered amplitudes of potentially different theories, called \emph{single copies}, and $\mathcal{M}_n$ is the \emph{double-copy} amplitude.
Due to the Kleiss--Kuijf (KK)~\cite{KLEISS1989616} and BCJ~\cite{Bern:2008qj} relations, the number of independent color-ordered amplitudes forming a \emph{BCJ basis} is $(n-3)!$.
The indices $\alpha,\beta$ in \autoref{eq:KLT} refer to the color-orderings of the single-copy amplitudes and the sums run over the elements of any two BCJ bases, while the KK and BCJ relations ensure that the double-copy amplitude does not depend on the chosen bases.

The multiplication of the single copies is governed by $S_n$, the KLT kernel, which is a scalar function of Lorentz invariants.
Its form depends on the BCJ bases considered in the sum.
The kernel plays a crucial role in ensuring that the resulting {double-copy} amplitude has a healthy analytical structure.
It cancels poles that are present in both ordered amplitudes, to prevent double poles in $\mathcal{M}_n$, and provides missing poles so that all physical factorization channels are generated.
This requires the KLT kernel to have a precise structure, which was found to be closely related to the amplitudes of the bi-adjoint scalar theory (BAS)~\cite{Cachazo:2013iea}. 

The BAS Lagrangian is given by
\begin{align}
\label{eq:BASLag}
\mathcal{L}_{\rm BAS} = -\frac{1}{2}(\partial_{\mu}\phi^{a\tilde a})^2 - \frac{g\bas}{6}f^{abc}\tilde{f}^{\tilde a \tilde b \tilde c}\phi^{a \tilde a}\phi^{b\tilde b}\phi^{c\tilde c}\, ,
\end{align}
where the scalar field has two adjoint color-group indices.\footnote{We refer to these indices as color ones although they are associated to global symmetries.}
We normalize the adjoint generators and structure constants such that $[T^{a},T^{b}]= f^{abc}T^c$ and $\Tr\!\left(T^aT^b\right)=\delta^{ab}$. 
Using the decomposition of the structure constants, the full bi-adjoint $n$-point amplitude can be written in terms of linearly independent traces of the generators,
\begin{align}
\mathcal{A}_n^{\textsc{bas}} = 
\sum_{\alpha, \beta \in S_{n-1}} 
{\rm Tr}(T^{a_{\alpha_1}}T^{a_{\alpha_2}}\cdots T^{a_{\alpha_n}})
\; m_n[\alpha|\beta]
\; {\rm Tr}(\tilde{T}^{\tilde{a}_{\beta_1}}\tilde{T}^{\tilde{a}_{\beta_2}}\cdots \tilde{T}^{\tilde{a}_{\beta_n}})
\,.
\label{traceDecomposition}
\end{align}
The objects $m_n[\alpha|\beta]$ are called doubly color-ordered amplitudes. 
They can be computed by summing over the trivalent graphs that contribute to the color orderings of both arguments, with appropriate relative signs.
For example, at 3-point we have $m_3[123|123]=g\bas$.

At 4-point, it is useful to define the 
vector of color factors,
	\begin{equation}\label{eq:colorsBAS}
			{c}\zero= \Big((1234),(1243),(1324),(1342),(1423),(1432)\Big)^\transpose   \,,
	\end{equation}
where $(1234)\equiv\tr(T^{a_1}T^{a_2}T^{a_3}T^{a_4})$, etc.
We define ${\tilde c\zero}$ similarly as $c\zero$ for the color factors with tilded indices.
The full BAS amplitude is then written compactly in matrix form as,%
\footnote{We will often omit vector arrows and transposes, writing for instance $c\zero \cdot {\bf m}_4 \cdot {\tilde c}\zero$ instead of $\vec c\zero^{\,{\scaleto{\mathsf{T}}{3.5pt}}} \cdot {\bf m}_4 \cdot \vec {\tilde c}\zero$.
}
\begin{align}
 \mathcal{A}_4^{\textsc{bas}}
		= c\zero \cdot {\bf m}_4 \cdot {\tilde c}\zero\,,
\end{align}	
with
{ \allowdisplaybreaks
	\begin{align}
	{\bf m}_4 &\!=\!\!
\begin{pmatrix}
 m_4[1234|1234] &  m_4[1234|1243] &  \cdots &  m_4[1234|1432] \\
  m_4[1243|1234] &  m_4[1243|1243] & \cdots &  m_4[1243|1432] \\
  \vdots & \vdots & \ddots & \vdots\\
  m_4[1432|1234] &  m_4[1432|1243] &  \cdots &  m_4[1432|1432]
\end{pmatrix} 
\label{eq:matrixm4}
\!\!=\!
g\bas^2
\begin{pmatrix}
\frac{1}{s}+\frac{1}{u} &  -\frac{1}{s} &  \cdots &  \frac{1}{s}+\frac{1}{u} \\[0.5mm]
 -\frac{1}{s}&\ \frac{1}{s}+\frac{1}{t} & \cdots &  -\frac{1}{s} \\[0.5mm]
  \vdots & \vdots & \ddots & \vdots\\[0.5mm]
  \frac{1}{s}+\frac{1}{u} &  -\frac{1}{s} &  \cdots & \  \frac{1}{s}+\frac{1}{u}
\end{pmatrix}
. \nn
\end{align}
where we use the conventions $s=s_{12}$, $t=s_{13}$ and $u=s_{14}$ with $s_{ab}=(p_a+p_b)^2$ and all momenta incoming. 

Remarkably, the KLT kernel can be identified as the inverse matrix of doubly color-ordered amplitudes \cite{Cachazo:2013iea},
\beq \label{kernelisminv}
    S_n[\alpha|\beta]=(m_n[\alpha|\beta])^{-1} \ ,
\eeq
where the indices $\alpha$ and $\beta$ should be restricted to any two BCJ bases (in which {color-ordered} amplitudes are independent), such that the $m_n[\alpha|\beta]$ sub-matrix is of full rank and can be inverted.
The rows and columns of ${\bf m}_n$ all satisfy the KK and BCJ relations,
and the double copies of BAS amplitudes are trivial:
\begin{equation}\label{eq:BASdoublecopy}
     m_n[\alpha|\delta]
     = \sum_{\beta, \gamma}^{(n-3)!} \,
     m_n[\alpha|\beta]\,
     S_n[\beta | \gamma] \, 
     m_n[\gamma|\delta]\, , 
\end{equation}
where the uncontracted $\alpha,\delta$ indices correspond to the color orderings that are left untouched in this relation.

The BAS {theory} can also be double-copied with another single-copy theory, in which case the KLT product encodes the KK and BCJ relations.
This is simple to illustrate at 4-point, where $m_4[\alpha|\beta]$ has rank $1$ and the kernel is simply the inverse of a number for fixed $\alpha,\beta$.
In this case, \autoref{kernelisminv} implies that there are choices of BCJ bases that render the KLT relation trivial, such as in
    \beq 
	A_{4}[\alpha] = m_4[\alpha|\beta]\;S_4[\beta|\gamma]\;A_{4}[\gamma]\,\quad {\rm or} \quad  A_{4}[\alpha] = 
	A_4[\gamma]\;S_4[\gamma|\beta]\;m_{4}[\beta|\alpha]\,,
    \eeq 
when $\alpha=\gamma$.
However, because of the KK and BCJ relations, the left-hand side does not depend on the $\beta,\gamma$ bases chosen in the product.
For example, (note $\alpha \neq \gamma$, but arbitrary $\beta$)
\begin{align} \label{tradKKBCJ}
    A_4[1234] 
    &= m_4[1234|\beta]\; S_4[\beta|1243]\; A_4[1243] = \frac{t}{u}\, A_4[1243]\,,
\end{align}
is exactly a BCJ relation, while one of the KK relations is given by
\begin{align} \label{tradKK}
    A_4[1234] 
    &= m_4[1234|\beta]\, S_4[\beta|1432]\, A_4[1432]=  A_4[1432]\,.
\end{align}
It was emphasized in \cite{Chi:2021mio} that the BAS behaves like an identity element in the KLT product, which is why it is also referred to as the \emph{zeroth copy}.

The identification of the BAS as the identity element of the KLT product leads to generalizations of this product associated to modifications of the BAS theory. 
This is exactly the case for the field theory form of the KLT relation between the open- and closed- string amplitudes \cite{Mizera:2016jhj}.
However, not all modifications of the BAS theory result in acceptable KLT kernels.
It was found that these corrections should preserve the rank of the matrix of doubly color-ordered amplitudes, which is $(n-3)!$ \cite{Chi:2021mio}.
This is called the \textit{minimal rank condition}.
In this paper, we will focus on higher-derivative (h.d.)\! corrections suppressed by powers of an EFT cutoff scale $\Lambda$, i.e.\ ${\bf m}_n\hd = {\bf m}_n + \mathcal{O}(1/\Lambda)$.
In the decoupling limit, $\Lambda\rightarrow\infty$, one therefore recovers the traditional KLT product.

In the generalized KLT formalism, the kernel $S_n\hd$ is the inverse of a full-rank sub-matrix of ${\bf m}_n\hd$ and satisfies
\beq \label{gen1x1=1}
{ m}_n\hd[\alpha|\delta]= 
\sum_{\beta, \gamma}
    { m}_n\hd[\alpha|\beta]
\; S_n\hd[\beta|\gamma]
\; { m}_n\hd[\gamma|\delta] \,.
\eeq
The consistency conditions that follow from products of the identity element (BAS) with another theory are called the generalized KKBCJ relations.
For single-copy color-ordered amplitudes, $A\gen_{n,\textsc{l/r}}$, they have the form
\begin{align} 
    A\gen_{n,\textsc{r}}[\alpha] &= \sum_{\beta, \gamma}
    m_n\hd[\alpha|\beta]\;
    S_n\hd[\beta|\gamma]\; 
    A\gen_{n,\textsc{r}}[\gamma]\,,
    \nn\\
    A\gen_{n,\textsc{l}}[\alpha] &= 
    \sum_{\beta, \gamma}
    A\gen_{n,\textsc{l}}[\gamma]\;
    S_n\hd[\gamma|\beta]\;
    m_n\hd[\beta|\alpha]\,,
\end{align}
and allow to bootstrap the single-copy amplitudes $A\gen_{n,\textsc{l/r}}$ that can take part in the double copy,
    \beq 
    \label{eq:genKLT}
       \mathcal{M}_n\gen =
       \sum_{\alpha, \beta}
    A\gen_{n,\textsc{l}}[\alpha]\;
    S_n\hd[\alpha|\beta]\;
    A\gen_{n,\textsc{r}}[\beta]\,.
    \eeq
Depending on the form of ${
m}_n\hd$ and $S_n\hd$, the generalized KKBCJ relations for $A\gen_{n,\textsc{r}}$ and $A\gen_{n,\textsc{l}}$ may be different.
We emphasize that both $A_{n,\textsc{l}/\textsc{r}}$ and $A_{n,\textsc{l}/\textsc{r}}\gen$ may in principle contain higher-derivative corrections.
The prime indicates that the amplitudes satisfy generalized KKBCJ relations, which allow for more operators.

That \autoref{eq:genKLT} produces a healthy amplitude when ${\bf m}_n\hd$ has minimal rank is a non-trivial empirical result \cite{Chi:2021mio}. 
The generalized KLT formalism allows for a systematic study of the space of theories that can appear as input and output of the double-copy procedure.
Although the bootstrap equations strongly constrain the higher-derivative corrections that are allowed in the input (single-copy) amplitudes, the generalized KLT formalism allows for more independent operators in the single copies than its traditional version. 
However, up to the orders checked explicitly in~\cite{Chi:2021mio}, it was found that the space of generalized output (double-copy) amplitudes $\mathcal{M}_n\gen$ is the same as  $\mathcal{M}_n$.
At 4-point, where the double-copy relation contains a single term, `similarity transformations' were proposed in~\cite{Chi:2021mio} to explain this fact (see also \cite{elvangQCDmeetsGravity2021}).
We aim to shed further light on this observation in \autoref{sec:seeds} and  \autoref{s:doublecopystructure}. 

As an example, the 4-point amplitude of the BAS theory with higher-derivative corrections is
	\begin{equation}\label{eq:anyBAS}
 	\mathcal{A}_4^{\textsc{bas}\text{+h.d.}}
		=  c\zero \cdot {\bf m}_4\hd \cdot  {\tilde c}\zero \,.
	\end{equation}
Solving the minimal rank condition, the ${\bf m}_4\hd$ matrix of doubly color-ordered amplitudes corresponding to BAS+h.d.\ can be written as
 \cite{Chi:2021mio} 
 \beq \label{eq:matrixfs}
 {\bf m}_4\hd=
\left(
\begin{array}{cccccc}
 f_1(s,t) & f_2(s,t) & f_2(u,t) & f_2(s,t) & f_2(u,t) & f_1(s,t) \\
 f_2(s,u) & f_1(s,u) & f_2(t,u) & f_1(s,u) & f_2(t,u) & f_2(s,u) \\
 f_2(u,s) & f_2(t,s) & f_1(t,s) & f_2(t,s) & f_1(t,s) & f_2(u,s) \\
 f_2(s,u) & f_1(t,u) & f_2(t,u) & f_1(t,u) & f_2(t,u) & f_2(s,u) \\
 f_2(u,s) & f_2(t,s) & f_1(u,s) & f_2(t,s) & f_1(u,s) & f_2(u,s) \\
 f_1(s,t) & f_2(s,t) & f_2(u,t) & f_2(s,t) & f_2(u,t) & f_1(s,t) \\
\end{array}
\right)\,,
 \eeq
 where \begin{equation}
	\label{eq:f1constraint}
		f_1(s,t) = f_1(u,t) \equiv  \frac{f_2(s,t)f_2(u,s)}{f_2(t,s)}\,,
	\end{equation}
and $f_2$ satisfies the bootstrap equation
	\begin{equation}\label{eq:f2constraint}
		f_2(s,t)f_2(t,u)f_2(u,s) = f_2(t,s)f_2(s,u)f_2(u,t)\,.
	\end{equation}
Furthermore, the aforementioned assumption 
${\bf m}_4\hd = {\bf m}_4 +\mathcal{O}(g\bas^2/\Lambda^2)$
requires $f_2(s,t) = -g\bas^2/s+\mathcal{O}(g\bas^2/\Lambda^2)$.
Additional constraints on $f_2$ arise if we forbid extra particles in the BAS+h.d.\ theory.

For later reference, we note that the alternative ordering of the single traces, $c\zero = \big((1324),(1234),(1243),(1423),(1432),(1342)
\big)^\transpose$,
exposes a block matrix structure, 
	\begin{equation}
	\label{eq:m4block}
		{{\bf m}}_4\hd = \begin{pmatrix}
			\tilde{{\bf m}}_{4}\hd  & \, \tilde{{\bf m}}_{4}\hd \\[1.1mm]
			\tilde{{\bf m}}_{4}\hd & \, \tilde{{\bf m}}_{4}\hd
		\end{pmatrix}
\,, \quad {\rm with} \quad
	\tilde{{\bf m}}_{4}\hd \equiv
\left(
\begin{array}{ccc}
 f_1(t,s) & f_2(u,s) & f_2(t,s)  \\
 f_2(u,t) & f_1(s,t) & f_2(s,t)  \\
 f_2(t,u) & f_2(s,u) & f_1(s,u) 
\end{array}
\right)\,.
\end{equation}
This structure also exists if we turn off the higher-derivative corrections, with
\beq \label{m3x3}
\tilde{{\bf m}}_{4} = g\bas^2 \left(
\begin{array}{ccc}
 \frac{1}{t}+\frac{1}{u} & -\frac{1}{u} & -\frac{1}{t} \\[2mm]
 -\frac{1}{u} & \ \frac{1}{s}+\frac{1}{u} & -\frac{1}{s} \\[2mm]
 -\frac{1}{t} & -\frac{1}{s} & \ \frac{1}{s}+\frac{1}{t} \\[2mm]
\end{array}
\right)\,.
\eeq 
Restricting to the smaller $\tilde{{\bf m}}_{4}\hd$ and $\tilde{{\bf m}}_{4}$ matrices will prove useful in the following sections.
We will actually only use these hereafter and drop the tildes for convenience.
However, such a block matrix structure generally only exists at lowest derivative order for $n>4$ particles. 
The 4-point case is special because the kinematics is invariant under reversal of the particle labels:
$\tilde f(1,2,3,4)\equiv  f(s_{12},s_{13})=  f(s_{43},s_{42}) \equiv \tilde f(4,3,2,1)$, for any function $f$ of the Mandelstam invariants.

\subsection{The generalized numerators approach}\label{reviewCarrasco}

Another approach to the double copy is based on the color-kinematics (CK) duality \cite{Bern:2010ue,Bern:2008qj}.
The basic idea is to use the decomposition of an on-shell $n$-point amplitude $\cA_n$ on trivalent graphs $g$,
\beq
\cA_n=\sum_g\frac{c_g n_g}{d_g} \ ,
\label{CKamplitude}
\eeq
where $d_g$ is the product of the (inverse) propagators it involves;
$c_g$ traditionally correspond to color factors associated to that same graph (i.e.\ combinations of generators of the gauge algebra);
while $n_g$ are the kinematic numerators that depend on Lorentz invariants and possibly on polarization vectors.
Given an amplitude $\cA_n$, the numerators $n_g$ are not unique.
CK duality is then a property of amplitudes for which there exists a choice of numerators $n_g$ which verify the same algebraic relations as those of the color factors $c_g$, inherited from the gauge algebra.
In certain theories, such as Yang--Mills, all tree-level amplitudes satisfy the CK duality.
In particular, the \text{SU}(N) color factors of 4-point amplitudes  obey Jacobi identities of the form%
\footnote{The signs in the Jacobi relation are fixed by the $c_s,c_t,\,c_u$ conventions. We use $c_s=f^{abe}f^{ecd}$, ${c_t=f^{ace}f^{edb}}$ and $c_u=f^{ade}f^{ebc}$.}
\begin{equation}
        c_{g_a} + c_{g_b} + c_{g_c} = 0\,,
    \end{equation}
and antisymmetry relations upon interchanging two legs on one vertex in the $g_a,g_b,g_c$ graphs.
Any numerator satisfying these \emph{adjoint algebraic relations} will be called an \emph{adjoint numerator}.

In gauge theories, the color relations ensure the gauge invariance of the amplitude under shifts of the polarization vectors contained in $n_g$, $\mathcal{A}|_{\eps_i\to p_i}=0$, for any particle label $i$.
This implies, in turn, that $c_g$ can be replaced by any expression which satisfies the same relations without spoiling gauge invariance, and the latter applies in particular to another copy of a CK-dual $n_g$.
The BCJ double-copy procedure~\cite{Bern:2010ue,Bern:2008qj} is thus schematically
\beq \label{eq:BCJdoublecopy}
\cA_n=\sum_g\frac{c_gn_g}{d_g}\,,\quad
\tilde \cA_n=\sum_g\frac{c_g\tilde n_g}{d_g}
\quad\longrightarrow\quad
\mathcal{M}_n=\sum_g\frac{\tilde n_g n_g }{d_g} \,,
\eeq
where $\mathcal{M}_n$ is the double-copy amplitude. 
The numerator $\tilde n_g$ is not necessarily the same as $n_g$, but both are CK-dual to the same $c_g$ color factors.%
    \footnote{The CK duality has to be manifest in at least one numerator. Notice, however, that a manifest CK duality can be achieved through so-called generalized gauge transformations, which are not always trivial to perform.} 
For two copies of Yang--Mills, the resulting amplitude describes the scattering of gravitons (as well as scalars).

Let us illustrate this approach at 4-point with the well-known case of Yang--Mills theories.
There are three trivalent diagrams associated to the $s,t,u$ channels, and the amplitude reads
\beq
\cA_4^\textsc{ym}(1^+_a2^+_b3^-_c4^-_d)=
g_\textsc{ym}^2 \, 
\sqr{12}{}^2\angl{34}{}^2\(\frac{ f^{ace} f^{edb}}{st}-\frac{ f^{ade}f^{ebc}}{su}\)=\frac{c_sn_s}{s}+\frac{c_tn_t}{t}+\frac{c_un_u}{u}\,,
\eeq
where we defined
\beq
 c_\text{adj} \equiv \bmat c_s\\ c_t\\ c_u\emat\equiv\bmat  f^{abe} f^{ecd}\\  f^{ace} f^{edb}\\  f^{ade}f^{ebc}\emat \ , \quad  n \equiv \bmat n_s\\ n_t\\ n_u\emat
\equiv
    g_\textsc{ym}^2 \, \sqr{12}{}^2\angl{34}{}^2
\bmat 
\frac{1-2\alpha}{u}-\frac{\alpha}{t}\\[2mm] \frac{\alpha}{s}-\frac{1-2\alpha}{u}\\[2mm] \frac{\alpha}{t}-\frac{\alpha}{s}
\emat \ ,
\label{YMnums}
\eeq
with an arbitrary function $\alpha$.
Square and angle brackets denote spinor-helicity variables (for pedagogical introductions see e.g.\ \cite{Elvang:2013cua, Cheung:2017pzi}).
The Jacobi identity reads $c_s+c_t+c_u=0$ and the explicit formul\ae\ above allow to check that $n_s+n_t+n_u=0$ for any $\alpha$.
The color factors also satisfy antisymmetry relations such as $c_s|_{1\leftrightarrow 2}=c_s|_{a\leftrightarrow b} = -c_s$.
For $\alpha = 1/3$, the numerators are antisymmetric too: e.g.\ $n_s|_{1 \leftrightarrow 2} = n_s|_{t\leftrightarrow u} = -n_s$.
This shows the CK duality of the 4-point YM amplitude.
The double-copy method then leads to a diffeomorphism-invariant four-gravitons amplitude,
\beq
\mathcal{M}_4^\textsc{gr}
(1^{ +2}
2^{+2}
3^{ - 2}
4^{ - 2})
=\frac{n_s^2}{s}+\frac{n_t^2}{t}+\frac{n_u^2}{u} \ .
\eeq

The bi-adjoint scalar theory considered in the previous subsection also plays the role of a zeroth copy in the CK approach.
It has a (color-color) dual structure, 
\beq\label{bastrivalent}
    \cA_n^\text{\textsc{bas}}
    =g\bas^{n-2} \sum_g\frac{c_{g}\tilde c_{g}}{d_g} \ ,
\eeq
where the two color groups lead to $c_g$ and $\tilde c_g$.
Being both color factors in the adjoint representation, they verify Jacobi and antisymmetry relations and any of the two can be treated as a CK-dual numerator $n_g$.
Moreover, replacing $c_g$ in \autoref{eq:BCJdoublecopy} by one of the BAS color factors has a trivial effect.

Higher-derivative effects in the BAS theory can be included, while preserving the dual CK structure, by building generalized numerators, $c\hd(c,s_{ab})$
\cite{Carrasco:2019yyn, Carrasco:2021ptp} (see also \cite{Low:2020ubn}).
These verify the same adjoint algebraic relations as the $c_g$, but may depend on both the Mandelstam invariants $s_{ab}$ and color factors $c$.
The color factors are themselves not necessarily of adjoint type, but more generally built from products of traces of group generators.
Only their combination with Mandelstams is required to satisfy the adjoint algebraic relations.
In this paper, we focus on single traces and do for instance not consider generalizations in the form of double traces ({see e.g.\ \cite{Low:2020ubn}}).

Choosing $c\hd = c_g + \mathcal{O}(s_{ab}/\Lambda^2)$, the generalized numerators result in EFT amplitudes for the bi-adjoint scalar theory that retains the CK-duality,
\beq\label{gennumBAS}
\cA_n^{\textsc{bas}+\text{h.d.}}
=g\bas^{n-2}\sum_g
\frac{ c_g\hd \, {\tilde c}_g^{\, \text{h.d.}} } {d_g} \ .
\eeq
Similarly, higher-order corrections in a gauge theory that preserve the CK-duality can be obtained by replacing the color factors by generalized numerators,
\beq\label{eq:singlecopyCK}
\cA_n^{\textsc{ym}+\text{h.d.}}=
\sum_g\frac{ c_g\hd n_g}{d_g} \,,
\eeq
for unmodified $n_g$.
This can also be interpreted as the double copy between $\mathcal{A}_n^\textsc{ym}$ and $\cA_n^{\textsc{bas}+\text{h.d.}}$  amplitudes.
The gauge invariance of the amplitude is maintained because the $ c_g\hd$ satisfy the same relations as the $c_g$.
This is the approach taken in~\cite{Carrasco:2019yyn,Carrasco:2021ptp} to extend the CK double-copy method to higher-derivative EFT amplitudes.

To find the most general allowed numerators, one could construct an ansatz for ${c}\hd$ at each order in the Mandelstams and impose the adjoint algebraic relations.
Exploiting the underlying structure instead, \cite{Carrasco:2019yyn} demonstrated that all purely kinematic adjoint 4-point numerators can be built using a simple composition rule acting on existing lower-order adjoint kinematic numerators $j$ and $k$ (vectors),
\beq
  n( j,  \,  k)
=\bmat j_tk_t-j_uk_u \\ j_uk_u-j_sk_s\\j_sk_s-j_tk_t\emat \,.
\label{compositionRules}
\eeq 
This requires only one building block made out of kinematic invariants,
\beq\label{eq:basicbb}
 n^{(ss)}=\bmat t-u \\ u-s\\s-t \emat \, .
\eeq
Furthermore, all adjoint 4-point numerators involving one factor of color can be generated using \autoref{compositionRules} with one of the two input numerators containing color, and the additional rule
\beq 
   c \, ( j,\, d)= d \,  j \ ,
    \label{compositiond}
\eeq 
where $d$ is the color factor that is fully symmetric under external particle permutations,
\begin{align}
\label{eq:d}
d^{(abcd)}\equiv \frac{1}{3!}\sum_{\sigma\in S_3}\Tr(T^aT^{\sigma(b)}T^{\sigma(c)}T^{\sigma(d)})\,.
\end{align}
The only other color structures required as primary building blocks are the adjoint ones, $c_s= f^{abe} f^{ecd}$, etc.\ encountered in \autoref{YMnums}.
All possible adjoint structures at 4-point can then be obtained by successive applications of these composition rules, and linear combinations of these yield the most general $c\hd$~\cite{Carrasco:2019yyn}.
At 5-point, the situation is complicated by the presence of more composition rules and algebraic structures \cite{Carrasco:2021ptp}.

In the next section, instead of constructing adjoint numerators from lower-order adjoint numerators and applying composition rules, we will build all of them from simpler \emph{non-adjoint} objects. 
This alternative construction procedure can be extended to higher multiplicities without much complication. 

\section{Numerator construction from seeds at any multiplicity}
\label{sec:seeds}

We now propose an alternative method to construct generalized adjoint numerators of the $\text{SU}(N)$ (or $\text{U}(N)$) unitary group.
At this stage, we are only interested in the adjoint numerators themselves, without regard to the factorization properties and particle content of amplitudes in which they enter. 
\hyperref[s:doublecopystructure]{Section\,\ref*{s:doublecopystructure}} discusses which extra constraints are imposed by such considerations.
The current section applies to any number of particles, while the construction is repeated explicitly in the next section at 4-point.

\subsection{Numerator seeds}

The main observation is that the adjoint color factor, $c_\text{adj}$, consisting of products of the structure constants, can be written in terms of linear combinations of single traces of the group generators (see also \cite{Naculich:2014rta}),
    \beq \label{cadjisJc0}
        c_\text{adj} = {\bf J}\cdot c\zero\,,
    \eeq 
where 
    \beq \label{eq:singletracen}
        {c}\zero = \big(
        (123...n), 
        ~
        \text{ all permutations of}~ \{2,3,...,n\}
        \big)^\transpose\,,
    \eeq
The matrix ${\bf J}$ contains only $\{\pm 1,0\}$ entries and will play a central role in our construction.
Its entries are determined by decomposing the structure constants in terms of traces through
    \beq 
        f^{abc} = (abc) - (acb)\,.
    \eeq
Products of traces can then be combined using the $\text{SU}(N)$ completeness relation,
    \beq 
        \sum_a T^a_{ij}\,T^a_{kl} = \delta_{il}\delta_{kj} - \frac{1}{N}\delta_{ij}\delta_{kl}\,,
    \eeq
and one can show that the $1/N$ terms cancel for products of structure constants.

In this way, the matrix ${\bf J}$ relates the simple algebraic structure of single traces (contained in the $c\zero$ vector) to the more involved adjoint algebraic properties. 
It encodes the Jacobi identities, and the antisymmetry relations which follow from the permutation properties of the single traces.
The matrix {\bf J} can be decomposed as follows:
\begin{align}
\label{eq:rankdecomp}
{\bf J} = {\bf A}\cdot {\bf B}\,
\end{align}
where ${\bf B}$ is the $(n-2)!\times (n-1)!$ matrix of rank $(n-2)!$ that relates $c\zero$ to the color factors in a Del Duca--Dixon--Maltoni (DDM) basis \cite{DelDuca:1999rs}, and ${\bf A}$ is the $(2n-5)!!\times (n-2)!$ matrix of rank $(n-2)!$ that relates the DDM basis to $c_\text{adj}$.
Conventions can also be chosen such that both $\bf A$ and $\bf B$ are sub-matrices of ${\bf J}$, see~\cite{Naculich:2014rta}.

Therefore, any object $n\zero$ that satisfies the same algebraic properties as the single traces will be mapped to an adjoint numerator under multiplication by ${\bf J}$.
The vector of single traces has linearly independent entries that are given in terms of permutations of one functional form, which is cyclically invariant in its arguments,
    \beq 
        (ab ... c) = (b ... ca)\,.
    \eeq 
In matrix notation, the algebraic properties can be summarized as follows.
Under a relabeling $\sigma$ of the particles, $c\zero$ transforms as 
\beq 
    c\zero \xrightarrow[\sigma]{} M_{c\zero,\sigma} \cdot c\zero\,,
\eeq 
where $M_{c\zero,\sigma}$ is a permutation matrix.
We shall call objects that obey the algebraic properties of the single traces,
\beq 
    n\zero \xrightarrow[\sigma]{} M_{c\zero,\sigma} \cdot n\zero\,,
\eeq 
\emph{numerator seeds}, or \emph{seeds} for short, as they can be used to generate adjoint numerators.
We shall use both $n\zero$ and $c\zero\hd$ to refer to numerator seeds.
The latter notation emphasizes that the cyclically invariant functional form of the seeds depends on both Mandelstam invariants and color factors.
Such seeds generate the generalized adjoint numerators discussed in \autoref{reviewCarrasco}.

In \autoref{proof}, we prove that \emph{any} adjoint numerator can be constructed from a numerator seed, i.e.
    \beq \label{seedeq}
        n_\text{adj} = {\bf J}\cdot n\zero\,.
    \eeq
Since the seeds are straightforwardly constructed, this provides an efficient way to explore the space of possible (generalized) adjoint numerators. A similar result was proven for kinematic numerators in renormalizable Yang--Mills theory using a different method \cite{Du:2013sha}.

A set of linearly independent numerator seeds generally maps to a redundant set of adjoint numerators, namely ${\bf J}\cdot n\zero = {\bf J}\cdot n\zero'$ could happen even for $n\zero \neq n\zero'$.
Therefore, identifying a set of seeds which generates independent adjoint numerators requires an extra step of reduction. 
This can be done by directly inspecting the general expression of $n_0$, or by relying on a construction which removes redundancies. We give explicit examples of the former below, while the latter can be achieved using the Moore--Penrose pseudo-inverse of ${\bf J}$, called ${\bf J}^+$ (see \autoref{proof}): it also follows from the argument of \autoref{proof} that ${\bf J}^+\cdot {\bf J} \cdot n\zero = {\bf J}^+\cdot {\bf J}\cdot n\zero'$ is a valid numerator seed.%
    \footnote{It is also interesting to note that the projection by ${\bf J}^+\cdot {\bf J}$ is equivalent to imposing the KK relations on the seeds. As shown in \cite{Naculich:2014rta}, a vector $\vec{v}$ satisfies the KK relations if and only if $\vec{r}_i\cdot \vec{v}=0$, where $r_i$ are the right null-vectors of ${\bf J}$. Since the right null-space of ${\bf J}$ is captured by $\mathbbm{1}-{\bf J}^+\cdot {\bf J}$, and $(\mathbbm{1}-{\bf J}^+\cdot {\bf J})\cdot {\bf J}^+\cdot {\bf J}\cdot n\zero =0$, the seed ${\bf J}^+\cdot {\bf J}\cdot n\zero$ satisfies the KK relations.
    }
A complete and independent set of numerator seeds can thus be obtained by projecting with ${\bf J}^+\cdot {\bf J}$ on all cyclically invariant functions.

\begin{table}[t]
    \centering
    \begin{tabular*}{\textwidth}{@{\extracolsep{\fill}}|c||*{15}{@{\;\:}c@{\;\:}|}}
         \hline
          {\bf k} 
          &\bf 1&\bf 2& \bf 3& \bf 4&\bf 5&\bf 6&\bf 7&\bf 8&\bf 9&
          \bf 10&\bf 11&\bf 12&\bf 13&\bf 14&\bf 15 
          \\\hline\hline
         {\bf 4-pt} 
         &1&1&1&2&2&2&3&3&3&4&4&4&5&5&5
         \\\hline
         {\bf 5-pt} 
         &0&0&1&2&5&8&14&21&32&45&63&84 &112 &144 &185 
         \\\hline
         {\bf 6-pt} 
         &1& 3& 9& 23& 54& 120& 243& 469& 861& 1509& 2546& 4158&$\cdots$&$\cdots$&$\cdots$
         \\\hline
    \end{tabular*}
    \caption{Number of independent scalar kinematic numerators at $\mathcal{O}(1/\Lambda^{2k})$ in the EFT expansion.
    The Gram determinant constraints relevant in 4 spacetime dimensions have been accounted for.
    The counting at 4- and 5-point was achieved in \cite{Carrasco:2019yyn} while the 6-point one is provided here for the first time.
    Although straightforward in principle, the numerically intensive reduction of the overcomplete set of numerators was not pushed beyond $k=12$ at 6-point.
   }
    \label{table6pt}
\end{table}

As an example of redundancies, permutation invariant functional forms result in valid numerator seeds, but they are mapped to zero and thus do not give rise to independent adjoint numerators.
In addition, ${\bf J}$ always combines seed entries with reversed ordering of particle labels, since ${c_\text{adj}\to (-1)^n c_\text{adj}}$ under reversal, at $n$-point.
The entries in a numerator seed can thus be ordered such that ${\bf J}$ has a block matrix structure, schematically:
${\bf J}_{a\times b} = \( {\bf J}_{a\times (b/2)}\,, \  (-1)^n {\bf J}_{a\times (b/2)}\)$. 
Therefore, the general numerator seed
    \beq 
    n\zero = \Big( 
        n\zero(1,2,...,n), \ ... \ ,
        n\zero(n,...,2,1), \ ... \ 
    \Big)\,,
    \eeq 
and the seed on which we impose (anti)symmetry under reversal on the functional form,
    \begin{align}
              \bar n\zero = \frac{1}{2}\Big( 
        &n\zero(1,2,...,n)+(-1)^n\,n\zero(n,...,2,1)
        , \ ... \ , \ \nn\\
        &\hspace{8mm}
        n\zero(n,...,2,1)+(-1)^n\,n\zero(1,2,...,n)
        , \ ... \ 
    \Big)  \,,
    \end{align} 
result in the same adjoint numerator.
At 4-point, these are the only sources of redundancy in the construction of adjoint numerators. 
There are further redundancies in the construction of CK-dual amplitudes, called generalized gauge transformations that will be addressed in section \autoref{s:gengauge}.
At higher multiplicity, redundancies can take a more complicated form, to be exemplified at 5-point in \autoref{5pts}. 

Even without identifying the specific algebraic origin of the redundancies, it is straightforward to just build the overcomplete set of seeds and identify numerically a basis of independent adjoint numerators. 
We provide the counting of the latter in \autoref{table6pt}, for numerators taking the form of polynomials of Mandelstam invariants. 
This table can be compared with \autoref{table-seed-counting} in \autoref{exnums}, listing the number of independent seeds, which grows faster with $n$ than the number of independent numerators.
The construction of adjoint numerators at 4-point agrees with the observation made in \cite{Carrasco:2019yyn}, namely that higher-order adjoint numerators can be obtained from lower order ones by multiplication with a permutation invariant function.
The results at 5-point also agree with the number of independent numerators listed in Table 2 of \cite{Carrasco:2021ptp} and we have constructed all the adjoint kinematic scalar numerators up to 6-point and $\mathcal{O}(1/\Lambda^{24})$.%
\footnote{We thank the authors of \cite{deNeeling:2022tsu} for private communications which lead us to correct our enumeration of 6-point numerators.}
We also provide explicit examples of numerator seeds and adjoint numerators of lowest orders in \autoref{exnums}. 

\subsection{Double copy and color-ordered amplitudes from numerators}\label{s:31}

The matrix ${\bf J}$ is also useful to obtain color-ordered amplitudes from adjoint numerators.
This approach was previously taken in \cite{Carrasco:2019qwr}, with a similar definition for ${\bf J}$.\footnote{
For us, ${\bf J}$ relates the adjoint color factors to the single traces while, in \cite{Carrasco:2019qwr}, they are instead related to a subset of adjoint color factors (which are independent under the Jacobi identities).}
For instance, writing the full bi-adjoint scalar amplitude of \autoref{bastrivalent} in matrix form and using the fact that $c_{\rm adj}$ and $c\zero$ are related by $\bf J$ through \autoref{cadjisJc0} yields
\beq 
    \cA^\text{\textsc{bas}} 
    = c_{\rm adj}
		\cdot 
		{\bf P}
		\cdot  {\tilde c}_{\rm adj}
	= c\zero \cdot {\bf J }^\transpose \cdot 
	{\bf P} \cdot {\bf J} \cdot 
	{\tilde c}\zero\,,
\eeq 
where ${\bf P}$ contains the propagators (and coupling constants) of the trivalent graphs on its diagonal. By definition, this produces the BAS matrix of doubly ordered amplitudes 
(see also \cite{Naculich:2014rta}), 
\beq 
    {\bf m} = {\bf J }^\transpose \cdot 
	{\bf P} \cdot {\bf J}\,.
\eeq 
Similarly, one can write the single-copy (color-ordered) amplitudes in terms of a numerator seed (replacing $\tilde c_\text{adj}$ by $n_\text{adj} = {\bf J} \cdot n\zzero{\textsc{r}}$),
\beq \label{nongensingle-0}
    A_\textsc{r} = 
	{\bf m}\cdot n\zzero{\textsc{r}}\,.
\eeq 
These single-copy amplitudes satisfy the traditional KK and BCJ relations as a consequence of the explicit factor of ${\bf m}$ and of the relation in \autoref{eq:BASdoublecopy}.
The double-copy amplitude can be obtained in the CK way,
    \beq \label{gravity}
        \mathcal{M}
        = 
        n_{\text{adj},\textsc{l}}\cdot 
        {\bf P}\cdot n_{\text{adj},\textsc{r}}
        =
        n\zzero{\textsc{l}} \cdot {\bf m} \cdot n\zzero{\textsc{r}}
        \,,
    \eeq 
or equivalently through the KLT relations,
    \begin{align}
        \mathcal{M} &= 
        \sum_{\alpha,\beta}^{(n-3)!}\,
        A_\textsc{l}[\alpha]\,
        S[\alpha|\beta]\,
        A_\textsc{r}[\beta]
        \nn\\
        &=
        \sum_{\alpha,\delta}^{(n-1)!}\,
        \sum_{\beta,\gamma}^{(n-3)!}\,
        n\zzero{\textsc{l}}[\alpha]\,
        m[\alpha|\beta]\,
        S[\beta|\gamma]\,
        m[\gamma|\delta]\,
        n\zzero{\textsc{r}}[\delta] 
        \nn\\
        &=n\zzero{\textsc{l}} \cdot {\bf m} \cdot n\zzero{\textsc{r}}\,.
    \end{align}
This exposes the special role played by the BAS matrix of color-ordered amplitudes to ensure the correct propagator structure of the double-copy amplitude.

The same method can be applied to obtain color-ordered amplitudes from generalized numerators. Defining a matrix ${\bf H}\hd$, which depends only on Lorentz invariants, one can decompose the numerator seeds (which, for simplicity, we build using only single traces) as
$c\hd\zero = {\bf H}\hd\cdot c\zero$\,.%
    \footnote{At 4-point, the form of the matrix ${\bf H}\hd$ will be derived in \autoref{4ptsCKSection}, where it will be shown to be diagonal. This is not generally true at higher multiplicity. 
    }
This allows \autoref{gennumBAS} to be rewritten as
    \beq \label{BAS+hd}
        \cA^{\textsc{bas}+\text{h.d.}} 
        = 
        c\zero\hd \cdot {\bf m} \cdot 
        \tilde c\zero^{\, \text{h.d.}}
        = 
        c\zero \cdot {\bf H}\hd_\textsc{r} \cdot {\bf m} \cdot 
        {\bf H}_\textsc{l}\hd \cdot \tilde c\zero\,.
    \eeq 
It follows that the higher-derivative color-ordered amplitudes can be constructed by left- and right-multiplication of the lowest order matrix ${\bf m}$,
    \beq \label{mhdleftright}
        {\bf m}\hd = 
        {\bf H}\hd_\textsc{r} \cdot {\bf m} \cdot 
        {\bf H}_\textsc{l}\hd\,.
    \eeq 
Similarly, starting from \autoref{eq:singlecopyCK}, one can write the higher-derivative single-copy (full) amplitude as 
\beq \label{fullsinglegen}
\cA_{\textsc{r}}\gen =c\hd\zero \cdot {\bf J }^\transpose \cdot 
	{\bf P} \cdot {\bf J}\cdot n\zzero{\textsc{r}} \ ,
\eeq
and color-ordered amplitudes as
\beq
A\gen_\textsc{r} =
{\bf H}\hd_\textsc{r} 
\cdot {\bf J }^\transpose \cdot 
	{\bf P} \cdot {\bf J}\cdot n\zzero{\textsc{r}} \ ,
\eeq
where we defined $n\zzero{\textsc{r}}$ such that $n_{\rm adj}={\bf J}\cdot n\zzero{\textsc{r}}$. 
Assuming that ${\bf H}\hd_\textsc{l}$ is of full rank, which always holds for an EFT expansion of the form ${\bf H}\hd_{\textsc{l}} = \mathbbm{1} + \mathcal{O}(s_{ab}/\Lambda^2)$, we can further define $n\gen\zzero{\textsc{r}}$ such that $n\zzero{\textsc{r}} \equiv {\bf H}_\textsc{l}\hd\cdot  n\gen\zzero{\textsc{r}}$, and we obtain
\beq  \label{numsinglecopy}
    A\gen_\textsc{r} 
	= {\bf H}\hd_\textsc{r} \cdot {\bf m} \cdot 
	{\bf H}\hd_\textsc{l}  \cdot 
	 n\gen\zzero{\textsc{r}}
	 = 
	{\bf m}\hd\cdot 
	 n\gen\zzero{\textsc{r}}
	\,.
\eeq 
Note that $ n\gen\zzero{\textsc{r}}$ also satisfies the properties of numerator seeds, and $A_\textsc{r}\gen$ satisfies the generalized KKBCJ relations.

The naive procedure for obtaining a double-copy amplitude in the generalized CK formalism is to replace $c\zero\hd$ in \autoref{fullsinglegen} by a kinematic seed $n\zzero{\textsc{l}}$ (or equivalently replace $c_\text{adj}\hd={\bf J}\cdot c\zero\hd$ by $n_{\text{adj},\textsc{l}}$).
This results in the same double-copy amplitude $\mathcal{M}$ as one could have been obtained without generalized numerators (\autoref{gravity}).
The same occurs in the generalized KLT formalism:
    \begin{align}
        \mathcal{M}\gen &=
        \sum_{\alpha,\beta}^{(n-3)!}\,
        A_\textsc{l}\gen[\alpha]\,
        S\hd[\alpha|\beta]\,
        A_\textsc{r}\gen [\beta]
     \nn\\
        &=\sum_{\alpha,\delta}^{(n-1)!}\,
        \sum_{\beta,\gamma}^{(n-3)!}\,
        {n}\gen\zzero{\textsc{r}}[\alpha]\,
        m\hd[\alpha|\beta]\,
        S\hd[\beta|\gamma]\,
        m\hd[\gamma|\delta]\,
         n\gen\zzero{\textsc{r}}[\delta]
        \nn\\
        &= 
         n\gen\zzero{\textsc{l}}\cdot
        {\bf m}\hd\cdot
        {n}\gen\zzero{\textsc{r}}
        =
        n\zzero{\textsc{l}}\cdot
        {\bf m}\cdot
        {n}\zzero{\textsc{r}} 
        = \mathcal{M}\,,
    \end{align}
where we distinguish sums over all $(n-1)!$ color factors and sums over BCJ bases. 

We have thus derived that the double-copy amplitudes obtained by the generalized KLT relations can equivalently be obtained through 
the traditional KLT double copy. 
However, there are two caveats to this statement. 
First,
we derived this statement assuming that ${\bf m}\hd$ is constructed via generalized numerators.
It is unclear whether it then reproduces all solutions to the KLT bootstrap.
In  \autoref{s:doublecopystructure}, we prove that this is the case at 4-point, and we have performed initial checks at 5-point presented in \autoref{5pts}.
Second, while the double-copy amplitude obtained by $A_{\textsc{l}/\textsc{r}}\gen$ and a generalized kernel
is the same as the one obtained by
$A_{\textsc{l}/\textsc{r}}= 
({\bf H}\hd_{\textsc{l}/\textsc{r}})^{-1}\cdot A_{\textsc{l}/\textsc{r}}\gen$ and a traditional kernel,
it is unclear whether $A_{\textsc{l}/\textsc{r}}$ are physical amplitudes and what is their particle content. 
In \autoref{s:doublecopystructure}, we show at 4-point that the assumption that ${\bf H}\hd_{\textsc{l}/\textsc{r}}$ does not affect the BAS particle content implies that this is also the case for $A_{\textsc{l}/\textsc{r}}$.

\section{Seeds and generalized numerators at 4-point}
\label{4ptsCKSection}

In this section, we will work out the 4-point construction of scalar adjoint numerators from their seeds.
This serves as an illustration of the method, and prepares for a comparison with the generalized KLT formalism in the next section.

A product of structure constants can be written in terms of single traces as follows,
\begin{align}
    f^{abx}f^{xcd} = \big[(abcd)+(dcba)\big]
    - \big[(abdc)+(cdba)\big].
\end{align}
As noted before, the traces appear together with their reversed ordering in this relation. 
The adjoint color numerator can thus be written compactly in terms of traces as
\beq \label{cadj4def}
 c_\text{adj}=
\bmat  f^{12x} f^{x34}\\  f^{13x} f^{x42}\\  f^{14x}f^{x23}\emat
=
\pmx{0&1&-1\\-1&0&1\\1&-1&0} \cdot 
\pmx{(1324)+		(4231)\\
			(1234)	+	(4321)\\
			(1243)+ (3421)} 
			= {\bf J}_4 \cdot c\zero
			\,.
\eeq
This defines the matrix ${\bf J}_4$ with $\{\pm 1,0\}$ entries, which encodes the Jacobi identity as the vanishing sum of its rows. 
Notice that we redefined the vector of traces $c\zero$ shown in \autoref{eq:colorsBAS}, combining traces and their reversed orderings such that
\begin{align}\label{eq1}
  c\zero = 
        \big( c\zero(1,3,2,4), \ 
         c\zero(1,2,3,4), \ 
          c\zero(1,2,4,3)\big)^\transpose,\quad 
    c\zero(a,b,c,d)=(ab cd)+(dcba)
\,.
\end{align}
We have conventionally chosen the ordering in the arguments of $c\zero$ entries such that the  $s,t,u$ Mandelstams are invariant under the cyclic permutation of the (1,3,2,4), (1,2,3,4), (1,2,4,3), respectively.
For instance, $s|_{1\to3\to2\to4} = s$.

The 4-point BAS amplitude can now be rewritten as
	\begin{align}
	\label{eq:BASadj}
	\mathcal{A}_4^{\textsc{bas}} &= 
		g^2\bas
		\left( \frac{c_s \tilde{c}_s}{s} + \frac{c_t \tilde{c}_t}{t} + \frac{c_u \tilde{c}_u}{u} \right)
		= 
		c_{\rm adj}
		\cdot 
		{\bf P}_4
		\cdot  {\tilde c}_{\rm adj}
		\nn\\
		&=
		c\zero \cdot {\bf J }_4^\transpose \cdot {\bf P}_4 \cdot {\bf J}_4 \cdot 
		{\tilde c}\zero = c\zero \cdot
		{\bf m}_4
		\cdot {\tilde c}\zero\,,
	\end{align}
with
\beq \label{P4}
 {\bf P}_4 \equiv	g^2\bas\,\begin{pmatrix}
			~\frac{1}{s}~&0&0\\ 0 &	~\frac{1}{t}~& 0\\0 & 0& 	~\frac{1}{u}~
		\end{pmatrix} 
		\,, 
\eeq 
and $\tilde {\bf m}_4$ defined in \autoref{m3x3}.

It is clear that ${\bf J}_4$ multiplying any numerator seed that satisfies the same algebraic relations%
    \footnote{Since the 4-point $c\zero$ of \autoref{eq1} is invariant under cyclic permutations and the reversal of its arguments, numerator seeds for example transform as
        ${n\zero \xrightarrow[\sigma]{} M_{c\zero,\sigma} \cdot n\zero}$,
    with ${M_{c\zero,\sigma}=\tiny\bmat 1&0&0\\0&0&1\\0&1&0\emat}$ for  
    ${\sigma = \{1{\to}3{\to}2{\to}4{\to}1\}}$
    and 
    ${M_{c\zero,\sigma}=\tiny\bmat 1&0&0\\0&1&0\\0&0&1\emat}$ 
    for ${\sigma = \{1\leftrightarrow4,2\leftrightarrow3\}}$.
    }
as $c\zero$
results in an adjoint numerator. 
However, it is not immediately clear that \emph{all} adjoint numerators can be constructed in this way. In the following, we prove that indeed the construction via numerators seeds leads to the complete set of adjoint numerators.   
For the general proof at any multiplicity, see \autoref{proof}.
First, notice that ${\bf J}_4^\transpose/3$ 
satisfies ${\bf J}_4\cdot {\bf J}_4^\transpose/3\cdot {\bf J}_4={\bf J}_4$,
and thus (since ${\bf J}_4$ encodes the Jacobi identities as the sum of its rows)
${\bf J}_4\cdot {\bf J}_4^\transpose/3\cdot n_{\text{adj}} 
=n_\text{adj}$
for any vector $n_\text{adj}$ that satisfies the Jacobi identities. 
Therefore, 
${\bf J}_4^\transpose\cdot n_{\text{adj}}/3$
is the pre-image of any $n_\text{adj}$. Importantly 
${\bf J}_4^\transpose\cdot n_{\text{adj}}/3$
is also a numerator seed:
    \begin{align}
        [{\bf J}_4^\transpose \cdot n_\text{adj}](1,3,2,4) &= -n_{\text{adj},t}+n_{\text{adj},u}
    \end{align} 
is invariant under cyclic permutations of its arguments and the reversal of their order, thanks to the algebraic properties of $n_\text{adj}$. 
This completes the proof.
As an example, consider multiplying the adjoint color factor by ${\bf J}_4^\transpose$,
    \begin{align} \label{exampleSeedPureColor4pts}
        [{\bf J}_4^\transpose \cdot c_\text{adj}](1,3,2,4) = -f^{13x}f^{x42}+f^{14x}f^{x23} =3\[(1324)+(4231)\]-\sum_{\sigma\in S_3}(1\sigma(234))\,.
    \end{align} 
The resulting seed is equivalent to the usual $c\zero$ up to the addition of a fully permutation-invariant quantity (mapped to zero by ${\bf J}_4$), as it should since they generate the same adjoint color numerator.

It is illustrative to compare the construction via numerator seeds with the composition method to construct adjoint numerators \cite{Carrasco:2019yyn}, reviewed in \autoref{reviewCarrasco}. 
Both the composition rule of \autoref{compositionRules} and the basic kinematic building block of \autoref{eq:basicbb} can be rewritten as a numerator seed multiplied by ${\bf J}_4$, 
\begin{equation}\label{eq:Jcomposition}
		n(j, k) =  {\bf J}_4 \cdot \pmx{j_s\,k_s\\j_t\,k_t\\j_u\,k_u}, 
		\qquad
		{n}^{(ss)} = {\bf J}_4 \cdot \pmx{s \\t \\ u}.
	\end{equation}
The composition rule of \autoref{compositiond} can similarly be rewritten,
\begin{equation}\label{eq:Jcomposition1}
	 n( j, d) = 	\vec n({\bf J}_4\cdot j\zero\,, \ d) = {\bf J}_4\cdot 
		\left(d\,j\zero \right) \,,
\end{equation}
where $j\zero$ is in fact a numerator seed (vector).
This last composition rule encodes the simple statement that a fully symmetric object, such as $d$, takes an adjoint numerator to another one.
This means that, at 4-point, there is a direct correspondence between the construction of adjoint numerators from seeds and by composition.
This is however not the case at 5-point, where the composition rules~\cite{Carrasco:2021ptp} are not equivalent to a simple multiplication by ${\bf J}_5$.

\subsection{Kinematic numerator seeds}

Analogous to \autoref{eq1}, purely kinematic  numerator seeds for a scalar theory have the form 
\begin{align}\label{eq1-2}
    n\zero = \big(
        n\zero(1,3,2,4), \ 
        n\zero(1,2,3,4), \ 
        n\zero(1,2,4,3)
        \big)^\transpose,
\qquad 
n\zero(a,b,c,d) = g(s_{ac},s_{ab})\,,
\end{align}
where $g$ is a function of the Mandelstam invariants $s_{ab}\equiv(p_a+p_b)^2$.
Invariance under reversal is automatic at 4-point, while cyclic invariance requires $g(s,t)=g(s,-s-t)=g(s,u)$. A general polynomial expansion of $g(s,t)$ can then be written as%
    \footnote{To see this, first express $g(s,t)$ in term of $t+u$ and $t-u$.
    The requirement that $g(s,t)=g(s,-s-t)=g(s,u)$ imposes a symmetry under $t\leftrightarrow u$ exchange which requires that $t-u$ only arises in even powers.
    However, since $(t-u)^2=s^2-4\,t\,u$, one concludes that $g(s,t)$ can be written as an expansion in powers of just $s$ and $t\,u$, as claimed.}
\begin{align}\label{expansiong}
    g(s,t)=\sum_{i,j=0}\frac{a_{i,j} \, s^i \,  \big(t \, [-s-t]\big)^j}{\Lambda^{2i+4j}} 
    =a_{0,0}+\frac{a_{1,0}s}{\Lambda^2}+\frac{a_{2,0}s^2+a_{0,1}tu}{\Lambda^4}+\frac{a_{3,0}s^3+a_{1,1}stu}{\Lambda^6}+\cdots
    \,.
\end{align}
Consistent factorization on the poles and assumptions on the particle spectrum of a theory can impose further restrictions on the $a_{i,j}$ coefficients.
From this numerator seed, a single-copy scalar amplitude can be constructed following \autoref{nongensingle-0}, which leads to
\beq \label{nongensingle}
    A_\textsc{r} = 
	{\bf m}_4\cdot n\zzero{\textsc{r}}
	=         -g\bas^2\,
        \Big(s\,g(s,t)+t\,g(t,s)+u\,g(u,s)
        \Big)
        \pmx{1/tu \\
             1/us \\
            1/st }
	\,,
\eeq 
where we note that the function $g(s,t)$ only appears through a permutation-invariant overall factor 
(so that it does not affect the traditional KK and BCJ relations).

\subsection{Generalized numerator seeds}\label{s:coloredseeds}

At zeroth order in Mandelstam invariants, one can verify that there is only one linear combination of single traces ($c_\text{adj}$ defined in \autoref{cadj4def}) that satisfies the adjoint algebraic properties.
Therefore, the only necessary numerator seed containing only color information is given by the $c\zero$ vector defined in \autoref{eq1}. 
At this order, any other seed is related to $c\zero$ by the addition of a permutation invariant combination of traces.
Such seeds also map to $c_\text{adj}$ because permutation invariant combination of traces map to zero under multiplication by ${\bf J}$.

At higher orders in the kinematics, the most general functional form that is cyclically and reversal invariant is 
  \begin{align} 
		\label{eq:mostgengennum2}
        \bar {c}\zero^{\, \text{h.d.}}(1,2,3,4)
        &= 	
 		g(t,s)\,c\zero(1,2,3,4)
		+ 
		 	h(u,t)\,c\zero(1,2,4,3) + 
		    h(s,t) \,c\zero(1,3,2,4)\,,
    \end{align}
from which one defines the generalized seed vector
\begin{align}
\label{general4PtSeed}
  \bar c\zero^{\, \text{h.d.}} = 
        \big( \bar c\zero\hd(1,3,2,4), \ 
         \bar c\zero\hd(1,2,3,4), \ 
          \bar c\hd(1,2,4,3)\big)^\transpose
\,,
\end{align}
analogously to \autoref{eq1}.
Here  $g(s,t)=g(s,-s-t)=g(s,u)$ which is the same constraint as before (\autoref{eq1-2}), and $h(s,t)$ is a priori a general function. Notice that
it is not trivial that we can write the equation above in terms of the vector $c\zero$ instead of the single traces separately.
This is a feature of the 4-point kinematics, which is invariant under reversal of particle labels: $ f'(1,2,3,4)\equiv  f(s_{12},s_{13})=  f(s_{43},s_{42}) \equiv  f'(4,3,2,1)$, for any function $f$ of the Mandelstam invariants.%
    \footnote{\label{foot:reversal-sum}%
    At higher multiplicity $n$, this is not generally true.
    While generalized numerator seeds can always be organized in terms of an $(n-1)!/2$ dimensional vector, factorizing out the color from the kinematic dependence generally requires all $(n-1)!$ single traces separately.
    }

While $\bar c\zero\hd$ is the most general numerator seed, for the purpose of constructing independent adjoint numerators, the $g(t,s)\,c\zero(1,2,3,4)$ term is redundant.
It can be canceled by adding a permutation invariant function and redefining the arbitrary $h(s,t)$. 
This means that we can restrict to the numerator seed
    \beq \label{seedmatrix}
        \bar c\zero^{\, \text{h.d.}}
        = \begin{pmatrix} 0&h(t,s)&h(u,s)\\
              h(s,t)&0&h(u,t)\\
              h(s,u)&h(t,u)&0
        \end{pmatrix} \cdot c\zero \equiv {\bf H}\hd \cdot c\zero\,,
    \eeq 
and still generate all possible adjoint numerators with color. We could have reached the same conclusion regarding the fact that the function $g$ can be absorbed in $h$ using the systematic algorithm which makes use of ${\bf J}^+$ which we discussed in \autoref{sec:seeds}.

\subsection{Generalized gauge transformations}\label{s:gengauge}

Up to this point, we have considered numerators independently from the amplitudes they generate. 
There does exist a freedom to shift a numerator without affecting the amplitude, if the other numerators they multiply satisfy Jacobi identities.
For instance, at 4-point, the redefinition
$n_{s}\to n_{s}+s\,\Delta, \ 
n_{t}\to n_{t}+t\,\Delta, \ 
n_{u}\to n_{u}+u\,\Delta$, for any function $\Delta$, results in
\beq \label{gengaugesum}
    \mathcal{A}_4 = \frac{ c_{s}\, n_{s}}{s} + 
        \frac{ c_{t}\, n_{t}}{t} + 
        \frac{ c_{u}\, n_{u}}{u} 
    \to 
    \mathcal{A}_4+(c_{s}+c_{t}+c_{u})\Delta\,,
\eeq 
which is just $\mathcal{A}_4$ if the color vector $\vec{c}$ satisfies the Jacobi identity.
In matrix notation, any shifts in the vector $\vec{n}$ proportional to
$(s,t,u)^\transpose$
leave the amplitude 
${\mathcal{A}={c}\cdot {\bf P}_4 \cdot {n}}$ 
invariant. Here $(s,t,u)^\transpose$ is the null vector of 
${\bf J}_4^\transpose\cdot{\bf P}_4$, where ${\bf J}_4^\transpose$ arises if $\vec{c}$ satisfies the Jacobi identity.
Such shifts are called \textit{generalized gauge transformations} because an actual gauge transformation, $\eps_i \to \eps_i+p_i$ for any particle label $i$, results in a similar vanishing shift of the amplitude.
Nevertheless, generalized gauge transformations are also present in non-gauge theories.

At the level of the numerator seeds, the generalized gauge transformations allow for shifts proportional to the null-vectors of 
$\tilde {\bf m}_4 = {\bf J}_4^\transpose \cdot {\bf P}_4 \cdot {\bf J}_4$, which are $(u,0,-s)^\transpose$ and $(t,-s,0)^\transpose$.
This includes the permutation invariant shift proportional to $(1,1,1)^\transpose$ (the null-vector of ${\bf J}_4$) that was used before and does not affect the constructed adjoint numerator.
Other shifts are possible that change the permutation properties of the seed and, in turn, may correspond to a non-adjoint numerator.
A particular generalized gauge transformation, given by 
    \beq
        \bar c\zero^{\, \text{h.d.}}
        \to 
        c\zero\hd = 
        \bar c\zero^{\, \text{h.d.}} + 
        \begin{pmatrix} 
        \frac{t\,h(s,t)+u\,h(s,u)}{s}       &-h(t,s)    &-h(u,s)\\[1mm]
        -h(s,t)  &   \ \, \frac{u\,h(t,u)+s\,h(t,s)}{t}        &-h(u,t)\\[1mm]
        -h(s,u)  &-h(t,u)     &\ \, \frac{s\,h(u,s)+t\,h(u,t)}{u} 
        \end{pmatrix} \cdot c\zero \,,
    \eeq 
has the property that the shift is itself a numerator seed and, therefore, maps into a valid seed $c\zero\hd$. 
Hence, to capture all amplitudes in a CK-dual theory, $\bar c\zero^{\, \text{h.d.}}$ in \autoref{seedmatrix} can be replaced by
\beq \label{Gmatrix}
c\zero\hd = \pmx{g(s,t)&0&0\\0&g(t,s)&0\\0&0&g(u,s)}\cdot c\zero 
\equiv {\bf G}\hd \cdot c\zero,
\eeq 
with {the redefinition}
\beq \label{poleing}
    g(s,t) = g(s,u) = 
    \frac{t \, h(s,t)+ u \, h(s,u)}{s}\,.
\eeq 
This defines the diagonal matrix ${\bf G}\hd$ for future reference.
We thus find that one can restrict to the functional form
   \begin{align} \label{minimalseed}
        {c}\zero\hd(1,2,3,4)
        &= 	
        g(t,s)\,c\zero(1,2,3,4)
    \end{align}
(if $g(s,t)$ is allowed to have simple poles).
We stress that this is not the most general numerator seed, nor does it construct the most general adjoint numerator, but it constructs the most general amplitude.

\section{Seeds and the generalized KLT bootstrap at 4-point}
\label{s:doublecopystructure}

The constructive approach to generalized adjoint numerators presented in the previous section enables a straightforward comparison with the generalized KLT approach by explicitly building BAS+h.d.\  amplitudes. Restricting to 4-point in this section, we show that the generalized CK and KLT formalisms are equivalent, and we study what this means for the possible double-copy amplitudes.

The matrix ${\bf m}\hd_4$ has the general form shown in \autoref{mhdleftright} but generalized gauge transformations allow us to write
    \begin{align}
        {\bf m}\hd_4
        &= 
        {\bf G}\hd_\textsc{r} \cdot {\bf m}_4 \cdot 
        {\bf G}_\textsc{l}\hd \nn \\[1mm] \label{mhd}
        &=
		g^2\bas\,
		\left( \begin{array}{ccc}
		 -\frac{s \, g_\textsc{r}(s,t) \, g_\textsc{l}(s,t)}{t \, u} & -\frac{g_\textsc{r}(s,t)\, 
		   g_\textsc{l}(t,s)}{u} & -\frac{g_\textsc{r}(s,t) \, g_\textsc{l}(u,s)}{t} \\[2mm]
		 -\frac{g_\textsc{r}(t,s) \, g_\textsc{l}(s,t)}{u} & -\frac{t \, g_\textsc{r}(t,s) \,
		   g_\textsc{l}(t,s)}{s \, u} & -\frac{g_\textsc{r}(t,s) \, g_\textsc{l}(u,s)}{s} \\[2mm]
		 -\frac{g_\textsc{r}(u,s) \, g_\textsc{l}(s,t)}{t} & -\frac{g_\textsc{r}(u,s) \,
		   g_\textsc{l}(t,s)}{s} & -\frac{u \, g_\textsc{r}(u,s) \, g_\textsc{l}(u,s)}{s \, t} \\[2mm]
		\end{array}
		\right)
		\,,
    \end{align}
where ${\bf G}_{\textsc{l}/\textsc{r}}$ is defined as in \autoref{Gmatrix} with $g_{\textsc{l}/\textsc{r}}(s,t)$ instead of $g(s,t)$.
In fact, this corresponds to a generalized KLT matrix showed in \autoref{eq:m4block}, 
with the solution to the minimal-rank bootstrap \hyperref[eq:f1constraint]{Eqs.\,(\ref{eq:f1constraint},\,}\hyperref[eq:f2constraint]{\ref{eq:f2constraint})} given by
    \begin{equation}\label{eq:f2eqgRgL}
        f_2(s,t) = -\frac{g^2\bas}{s}\, g_\textsc{r}(t,s)\,g_\textsc{l}(u,s) \,.
    \end{equation}
Another way to see that ${\bf m}\hd_4$ has minimal rank, is that the diagonal matrices ${\bf G}\hd_{\textsc{l}/\textsc{r}}$ have full rank and therefore preserve the rank of the matrix 
${\bf m}_4$.
Conversely, the relation above can be inverted to express $g_{\textsc l/\textsc r}$ in terms of $f_2$ (not uniquely), which means that they encompass any EFT solution $f_2$ of the bootstrap equations.
One can for instance take
	\begin{align}\label{eq:bootstrapsol}
		g_\textsc{r}(s,t) = \left( \frac{f(t,s)\,f(u,s)^2\,f(t,u)}{f(s,u)}\right)^{1/3}, \qquad
		g_\textsc{l}(s,t) = \left( \frac{f(t,u)\, f(u,t)}{f(s,t)\,f(s,u)} \right)^{1/3}
		\,,
	\end{align}
with $f(s,t) \equiv -s\,f_2(s,t)/g\bas^2$.
Thanks to the bootstrap condition for $f_2$, these satisfy the constraint $g(s,t)=g(s,-s-t)=g(s,u)$, as required for numerator seeds.
This shows that the generalized numerator seed of \autoref{minimalseed} generates any matrix of doubly ordered amplitudes that appears in the generalized KLT formalism.
The choice above is not unique since ${\bf m}\hd_4$ only depends on the product of $g_\textsc{l}$ and $g_\textsc{r}$.

To illustrate how \autoref{eq:bootstrapsol} works in an EFT expansion, let us consider the lowest-order terms in the bootstrap solution for a pure scalar theory \cite{Chi:2021mio},
\beq \label{bootstrapexp}
    \frac{f_2(s,t)}{g^2\bas} = -\frac{1}{s} + \frac{1}{\Lambda^4}(a_{1,0}\, t + a_{1,1}\, s) +
    \frac{a_{2,0}}{\Lambda^6}\,t\,(s+t) + \mathcal{O}\left(\frac{s_{ab}^3}{\Lambda^8}\right)\,,
\eeq
which determines the generalized matrix of doubly ordered amplitudes.
Besides the bootstrap equation (\autoref{eq:f2constraint}), further constraints have been imposed on this function, which ensure correct locality properties of the resulting BAS+h.d.\ theory.
From \autoref{eq:bootstrapsol}, it follows that the matrix determined by $f_2$ is equivalently obtained through \autoref{mhd} 
with
\begin{align}
    g_\textsc{r}(s,t) &= 1 +
    \frac{4a_{1,1}-a_{1,0}}{3\Lambda^4}\, 
        t \, (-t-s)
    + \frac{a_{1,0}-a_{1,1}}{3\Lambda^4}s^2 + \frac{a_{2,0}}{\Lambda^6}\,s\,t\,(-s-t)
    +\mathcal{O}\left(\frac{s_{ab}^4}{\Lambda^8}\right)\,,
    \nn\\[2mm] \label{solgfromf2}
    g_\textsc{l}(s,t) &= 1 - \frac{a_{1,0}-a_{1,1}}{3 \Lambda^4} (s^2 + 2\,t\,(-s-t))
    +\mathcal{O}\left(\frac{s_{ab}^4}{\Lambda^8}\right)
    \,.
\end{align}
These functions give rise to the numerator seeds and the associated adjoint numerators of the BAS+h.d.\ theory.

The minimal-rank bootstrap equations also allow for solutions that are not of the BAS+h.d.\ form.
However, modifications to the lowest-order 4-point kernel were found to increase the rank at higher multiplicities and lead to unhealthy double-copy structures \cite{Chi:2021mio}.

\subsection{Generalized single and double copies}

From \autoref{mhd}, it is now straightforward to obtain the generalized KLT kernel $S\hd_4$ in terms of the traditional BAS one, $S_4[\alpha|\beta] = 1/{ m}_4[\beta|\alpha]$\,:
    \beq \label{examplegenKernel}
        S_4\hd[\alpha|\beta] = 
        \frac{1}{{\bf G}_\textsc{l}\hd[\alpha|\alpha]}
            \, S_4[\alpha|\beta] \,
        \frac{1}{{\bf G}_\textsc{r}\hd[\beta|\beta]}\,.
    \eeq 
Due to the diagonal structure of ${\bf G}_{\textsc{l}/\textsc{r}}\hd$, each entry of the generalized kernel only depends on that of the BAS kernel with the same color ordering. 
Since the rank of both ${\bf m}_4$ and ${\bf m}\hd_4$ is one, this relation holds for any choice of single color orderings $\alpha$ and $\beta$.
One can explicitly check that this kernel obeys the KLT relation of \autoref{gen1x1=1},
    \begin{align}  
    { m}\hd_4[\alpha|\beta]
\; S\hd_4[\beta|\gamma]
\; { m}\hd_4[\gamma|\delta] 
&=
    {\bf G}\hd_\textsc{r}[\alpha|\alpha] \;
    { m}_4[\alpha|\beta]
\; S_4[\beta|\gamma]
\; { m}_4[\gamma|\delta] \;
    {\bf G}\hd_\textsc{l}[\delta|\delta] \nn\\
    &=
        {\bf G}\hd_\textsc{r}[\alpha|\alpha] \;
    { m}_4[\alpha|\delta] \;
    {\bf G}\hd_\textsc{l}[\delta|\delta] \nn\\
&= { m}\hd_4[\alpha|\delta]\,,
\end{align} 
and the generalized KKBCJ relations for color-ordered amplitudes are given in \autoref{numsinglecopy},
    \begin{align} \label{KKBCJgenEx}
        {\bf G}_\textsc{r}\hd[\alpha|\alpha]\;
        m_4[\alpha|\beta]\,S_4[\beta|\gamma]
        \;\frac{1}{{\bf G}_\textsc{r}\hd[\gamma|\gamma]}
        \;A\gen_\textsc{r}[\gamma]
        &= A\gen_\textsc{r}[\alpha]\,,\nn\\
        A\gen_\textsc{l}[\alpha]
        \;\frac{1}{{\bf G}_\textsc{l}\hd[\alpha|\alpha]}
        \;S_4[\alpha|\beta]\,m_4[\beta|\gamma]\;
        {\bf G}_\textsc{l}\hd[\gamma|\gamma]
                &= A\gen_\textsc{l}[\gamma]\,.
    \end{align}
For example, \autoref{tradKKBCJ} is generalized to
    \begin{align} \label{KKBCJpoles}
        A\gen_\textsc{r}[1234]
        &=
        \frac{t\, g_\textsc{r}(t,s)}
        {u\,g_\textsc{r}(u,s)}
        \,
        A\gen_\textsc{r}[1243]
        \,,
    \end{align}
which reduces to the traditional BCJ relation for $g_\textsc{r}(s,t)=1$.
It is worth remarking that the KKBCJ relations for $A\gen_\textsc{r}$ ($A\gen_\textsc{l}$) depend only on ${\bf G}_\textsc{r}\hd$ 
(${\bf G}_\textsc{l}\hd$).

Interestingly, the explicit form of the generalized KKBCJ relations points to an object
    \beq \label{genfromnongen}
        A_\textsc{r}[\alpha] \equiv 
        \frac{A\gen_\textsc{r}[\alpha]}
        {{\bf G}\hd_\textsc{r}[\alpha|\alpha]}\,,
    \eeq
which obeys the traditional KK and BCJ relations, and similarly for $A_\textsc{l}$. Although the notation suggests otherwise, $A_\textsc{r}$ may still contain higher-derivative corrections, but they are such that they do not affect {the form of the traditional} KK and BCJ relations.
If $A_\textsc{r}$ can be argued to be a valid amplitude, this could be an efficient method to construct generalized single copies. 
In addition, it would imply that the generalized KLT formalism does not 
lead to double copies with additional higher-derivative corrections besides the ones which can be obtained with the usual KLT kernel. Indeed,
    \beq\label{eq:gengravity}
        \dcamp = 
        A\gen_\textsc{l}[\alpha]\,
        S\hd_4[\alpha|\beta]\,
        A\gen_\textsc{r}[\beta]
        = 
        A_\textsc{l}[\alpha]\,
        S_4[\alpha|\beta]\,
        A_\textsc{r}[\beta]
        =
        \mathcal{M}
        \,,
    \eeq
where we stress that both primed and unprimed amplitudes may contain higher-derivative corrections and $S_4$ stands for the traditional BAS KLT kernel. 

However, as it stands, $A_\textsc{r}[\alpha]$ cannot be interpreted as a physical
amplitude, since it does not necessarily factorize properly on all channels, whose associated residues can be affected by ${\bf G}_\textsc{r}\hd$. 
We will  study the functional form of ${\bf G}_{\textsc{l}/\textsc{r}}\hd$, under the assumption of a fixed BAS particle content in \autoref{RestrictingParticles}.
This will lead to a slightly adapted but equivalent form for \autoref{genfromnongen}, such that $A_\textsc{r}[\alpha]$ and $A\gen_\textsc{r}[\alpha]$ have the same residues on all poles.

Leaving momentarily aside the question of physical residues, \autoref{genfromnongen} also provides a means to construct generalized single copies.
They can be obtained by multiplying amplitudes that satisfy the traditional KK and BCJ relations with ${\bf G}\hd_{\textsc{l}/\textsc{r}}$.
The same conclusion is reached in the numerator formalism (see \autoref{numsinglecopy}). Explicitly, comparing with \autoref{nongensingle}, we have that
\beq \label{rightamplitudesEX} 
        A_\textsc{r}\gen = 
        {\bf G}_\textsc{r}\hd \cdot 
        {\bf m}_4 \cdot 
        n\zero
        = 
        -g\bas^2\,
         \Big(s\,g(s,t)+t\,g(t,s)+u\,g(u,s)
        \Big)
        \pmx{g_\textsc{r}(s,t)/tu \\
             g_\textsc{r}(t,s)/us \\
             g_\textsc{r}(u,s)/st }\,,
    \eeq 
where $n\zero=\big(g(s,t),g(t,s),g(u,s)\big)^\transpose$.
All generalized amplitudes can be constructed in this way, showing that one can associate a generalized color numerator to any amplitude obtained from the generalized KKBCJ relations.
The first equality in \autoref{rightamplitudesEX}  also applies to gauge theories in which case $n\zero$ contains polarization vectors.
Imposing particular locality properties on this amplitude restricts the coefficients inside $g(s,t)$ and $g_\textsc{r}(s,t)$. We note that such constraints may be less restrictive than the constraints on $g_{\textsc{l}/\textsc{r}}$ coming from imposing a fixed particle content on ${\bf m}\hd_4$.
We will get back to this point in the following.

\subsection{Factorization properties and particle spectrum}
\label{RestrictingParticles}

So far, we have not been concerned with the particle content (and factorization properties) of the BAS+h.d.\ amplitudes constructed through \autoref{mhd}. 
Following \cite{Chi:2021mio}, we now impose that ${\bf m}\hd_4$ reduces to the BAS matrix ${\bf m}_4$ at lowest order and that the particle content of the theory is fixed to one bi-adjoint scalar.
This implies that ${f(s,t) = -s\,f_2(s,t)/g\bas^2}$ in \autoref{eq:bootstrapsol} is a polynomial of the form $ 1+\mathcal{O}(s_{ab}/\Lambda^2$).
Interpreting \autoref{eq:bootstrapsol} up to a fixed order in the $1/\Lambda$ expansion, the functions $g_{\textsc{l}/\textsc{r}}(s,t)$ are then also of the same form.

While double or spurious poles are avoided in the construction of ${\bf m}\hd_4$ with polynomial $g_{\textsc{l}/\textsc{r}}(s,t)$, the residues might be modified. 
Such modifications are either non-physical or can be interpreted as new particles appearing in the factorization channels. 
However, at 4-point, only contact-term higher-derivative corrections are allowed with a fixed single scalar particle content (since the 3-point amplitudes are not modified in the solution to the KLT bootstrap \cite{Chi:2021mio}). 
Imposing such conditions yields the following constraints:
\begin{equation}
\label{conditions}
\begin{gathered}
    m_4[1324|1324] = \frac{g_\textsc{r}(s,t) g_\textsc{l}(s,t)}{t} + 
    \frac{g_\textsc{r}(s,t) g_\textsc{l}(s,t)}{u}
    \underset{t\to 0}{\sim}\frac{1}{t}
    \implies 
    {g_\textsc{r}(s,0) g_\textsc{l}(s,0)}= 1
\,,\\
    m_4[1324|1234] =
    -\frac{g_\textsc{r}(s,t)
		   g_\textsc{l}(t,s)}{u}
    \underset{u\to 0}{\sim}-\frac{1}{u}
    \implies 
    {g_\textsc{r}(-t,t)
		   g_\textsc{l}(t,-t)} = 1\,,
\end{gathered}
\end{equation}
which enforce in particular the consistency conditions
\begin{equation}
g_{\textsc{l}/\textsc{r}}(s,0)=g_{\textsc{l}/\textsc{r}}(-s,0)\,,
\label{eq:consistency-condition}
\end{equation}
when using $g_{\textsc{l}/\textsc{r}}(s,t) = g_{\textsc{l}/\textsc{r}}(s, -s-t)$.
This implies that, on the $t$ and $u$ poles, the first variable of $g(s,t)$ necessarily appears in even powers.
The lowest order terms in the solutions are then
    \begin{align}
        g_\textsc{l}(s,t) &= 1+ \frac{{a_\textsc{l}}_{2,0}\,s^2+{a_\textsc{l}}_{0,1}\,t\,u}{\Lambda^4}
            +\frac{{a_\textsc{l}}_{1,1}\,s\,t\,u}{\Lambda^6}\,,
            \nn\\
        g_\textsc{r}(s,t) &= 1+ \frac{-{a_\textsc{l}}_{2,0}\,s^2+{a_\textsc{r}}_{0,1}\,t\,u}{\Lambda^4}
            +\frac{{a_\textsc{r}}_{1,1}\,s\,t\,u}{\Lambda^6}\,,
    \end{align}
where we emphasize that the same coefficient ${a_\textsc{l}}_{2,0}$ appears in both expansions.
These solutions are fully consistent with the expansions in \autoref{solgfromf2}, which were obtained from the bootstrap solution $f_2(s,t)$, assuming a fixed particle content \cite{Chi:2021mio}.\footnote{The consistency conditions also imply that  $f_2(s,t)$ does not contain even powers of just $s$.
Since $g_{\textsc{l}/\textsc{r}}(s,t) = g_{\textsc{l}/\textsc{r}}(s, -s-t)$ takes $f_2(s,0) \propto g_\textsc{r}(0,s)\,g_\textsc{l}(-s,s) / s$ to $g_\textsc{r}(0,-s)g_\textsc{l}(-s,0)/s$, the conditions of \autoref{eq:consistency-condition} indeed constrain $f_2(s,0)$ to be an odd function of $s$.}
With these solutions, one of the generalized KKBCJ relations, showed in \autoref{KKBCJpoles}, is given by
\beq 
        \left[
        1+\frac{({a_\textsc{l}}_{2,0}-{a_\textsc{r}}_{0,1})\:s(t-u)}{\Lambda^4}+\mathcal{O}\left(\frac{s_{ab}^4}{\Lambda^8}\right)
        \right]\,\frac{t}{u}\,A\gen_\textsc{r}[1243]
        = A\gen_\textsc{r}[1234]\,. 
\eeq

Now let us turn our attention to the object 
$        A_\textsc{r}[\alpha] \equiv 
        {A\gen_\textsc{r}[\alpha]}/
        {{\bf G}\hd_\textsc{r}[\alpha|\alpha]}
$
that appeared in \autoref{genfromnongen}.
Given that the functions $g_{\textsc{r}}$ in  ${{\bf G}\hd_\textsc{r}[\alpha|\alpha]}$ are polynomials, the $A_\textsc{r}[\alpha]$ and $A\gen_\textsc{r}[\alpha]$ functions have the same poles.
However, it is not guaranteed that the residues on these poles are consistent.
In particular, on any of the poles of $A_\textsc{r}[\alpha]$, the function ${{\bf G}\hd_\textsc{r}[\alpha|\alpha]}$ contributes a non-trivial inverse factor of $g_{\textsc{r}}(s,0)$ (or $g_{\textsc{r}}(t,0)$).
On the poles, these factors can also be obtained from the function
$g_{\textsc{r}}({\tiny \sqrt{{(s^2+t^2+u^2)}/{2}}},0)|_{t\to 0}=g_{\textsc{r}}(s,0)$, where the square root always appears in even powers in the Taylor expansion over the first variable thanks to the consistency condition of \autoref{eq:consistency-condition}.
Since $g_{\textsc{r}}({\tiny \sqrt{{(s^2+t^2+u^2)}/{2}}},0)$ is permutation invariant, we can redefine the amplitude 
$ A_\textsc{r}[\alpha]$ as
    \beq \label{gentonongenR}
    A_\textsc{r}[\alpha] \equiv \frac{g_\textsc{r}\! \(\sqrt{\frac{s^2+t^2+u^2}{2}},0\)}
    {{\bf G}_\textsc{r}\hd[\alpha|\alpha]}
    A\gen_\textsc{r}[\alpha]
    \eeq 
which still satisfies the traditional KK and BCJ relations while also having the same poles and residues as $ A\gen_\textsc{r}[\alpha]$.
Defining simultaneously 
    \beq \label{gentonongenL}
    A_\textsc{l}[\alpha] \equiv \frac{g_\textsc{l}\! \(\sqrt{\frac{s^2+t^2+u^2}{2}},0\)}
    {{\bf G}\hd_\textsc{l}[\alpha|\alpha]}
    A\gen_\textsc{l}[\alpha]\,,
    \eeq 
 it follows from \autoref{conditions} that 
$g_\textsc{l}\!\(\sqrt{{(s^2+t^2+u^2})/{2}},0\) = 1/g_\textsc{r}\!\(\sqrt{{(s^2+t^2+u^2})/{2}},0\)$,
which leads to the conclusion that the double-copy amplitude remains unchanged. 
In other words, there is an interplay between the left and right amplitudes and the generalized kernel that allows for the cancellation of any correction to the kernel, in a manner that does not affect the residues and poles of the amplitudes.

We therefore conclude that a double-copy amplitude obtained with a generalized kernel can equivalently be generated with the traditional BAS kernel (c.f.\ \autoref{eq:gengravity}). 
The single-copy amplitudes may then still include higher-derivative corrections, but only those that do not spoil the usual KK and BCJ relations.
To prove this statement, we assumed that no extra particles are added to the BAS spectrum. 
At 4-point, the generalization of the KLT formalism does thus not enlarge the space of possible double copies, but it does enlarge the space of single copies that can be used as input.

In our derivation, it is also clear that
\hyperref[gentonongenR]{Eqs~(\ref*{gentonongenR},}
\hyperref[gentonongenL]{\ref*{gentonongenL})}
can be used to obtain amplitudes that satisfy generalized KKBCJ relations ($A\gen_{\textsc{l}/\textsc{r}}$) from amplitudes satisfying the usual KK and BCJ relations 
($A_{\textsc{l}/\textsc{r}}$), with the same particle content. 
This, indeed, has exactly the same form in the generalized numerators approach shown in \autoref{rightamplitudesEX}.

\section{Results at 5-point}
\label{5pts}

Here, we use the numerator seeds described in \autoref{sec:seeds} to construct adjoint numerators at 5-point. 
We also discuss the seed redundancies, previously described at 4-point in \autoref{4ptsCKSection}.
The statement that the generalized KLT formalism does not enlarge the space of double-copy amplitudes had two caveats, see \autoref{s:31}. 
While these were fully addressed at 4-point in the previous sections, we only provide partial results at 5-point.
We check explicitly that the generalized numerators generate all the leading-order KLT kernels of \cite{Chi:2021mio}.
Furthermore, we study the factorization properties and the particle spectrum of the objects 
$A_{\textsc{l}/\textsc{r}}= 
({\bf H}\hd_{\textsc{l}/\textsc{r}})^{-1}\cdot A_{\textsc{l}/\textsc{r}}\gen$ for a restricted set of higher-derivative corrections.
For the corresponding generalized kernels, we achieve the same conclusion as at 4-point, namely that the double-copy amplitudes it produces can equally be obtained with the traditional BAS kernel and physical single-copy amplitudes. 

\subsection{Numerator seeds}

At 5-point, the kinematic numerator seeds have the functional form 
\begin{align}
\label{5ptkinseed}
    n\zero(1,2,3,4,5) = 
    \big( \,g\!\left(s_{12},s_{23},s_{34},s_{45},s_{51}\right) -
        g\!\left(s_{51},s_{45},s_{34},s_{23},s_{12}\right)\big)
        + \text{cyclic}\,,
\end{align}
where we have demanded invariance under cyclic permutations and antisymmetry under reversal.
Imposing this antisymmetry is however optional, as the components symmetric under reversal are mapped to zero by the ${\bf J}$-matrix.
We do not consider the parity-odd fully antisymmetric contraction of four independent momenta, which may lead to additional numerators.
The numerator seed vector is then given by
    \beq 
        {n}\zero = \big(n\zero(1,2,3,4,5),
                \quad 
        \text{all permutations of} ~\{2,3,...,n\}
        \big)^\transpose\,,
    \eeq
and all scalar adjoint numerators are easily built from $n_{\rm adj}= {\bf J}_5\cdot n\zero$.
The explicit form of ${\bf J}_5$ is given in \autoref{appJ5pt}. 
We chose a cyclic basis of Mandelstam invariants, which is possible at any multiplicity (see e.g.\ \cite{Boels:2016xhc}) and simplifies the particle permutation properties.

In addition, there are two independent adjoint numerators containing only color information. 
These are obtained from the following single-trace seeds,
\begin{equation}
\label{5ptc0s}
\begin{aligned}
    c{\zeroone} &= \big((12345),
               \quad 
      \text{all permutations of} ~\{2,3,...,n\}
        \big)^\transpose\,,
\\
      {c}\zerotwo &=  \big((1 3 5 2 4),
                \quad 
        \text{all permutations of} ~\{2,3,...,n\}
        \big)^\transpose\,.
\end{aligned}
\end{equation}
As before, it is not necessary to impose antisymmetry under reversal.
Any other allowed single-trace color numerator maps to a linear combination of ${\bf J}_5\cdot c\zeroone$ and ${\bf J}_5\cdot c\zerotwo$.
For instance, the two independent adjoint numerators $c_{\mathfrak{a},1,2}$ of \cite{Carrasco:2021ptp} are%
\footnote{The original definitions of these two objects are $c_{\mathfrak{a},1}(1,2,3,4,5) \equiv 
   f^{12x} f^{x3y} f^{y45}$ and  $c_{\mathfrak{a},2}(1,2,3,4,5) \equiv 
   2\,d^{123x}f^{x45}
    +d^{124x}f^{x35}
    -d^{125x}f^{x34}
    +2\,d^{234x}f^{x15}
    -2\,d^{235x}f^{x14}$~\cite{Carrasco:2021ptp}.}
\begin{align}
    c_{\mathfrak{a},1} = {\bf J}_5 \cdot c\zeroone, \qquad  c_{\mathfrak{a},2} =\frac{1}{6} \,{\bf J}_5\cdot (3c\zerotwo - c\zeroone)\,.
\end{align}

Besides numerator seeds featuring only kinematic and color information, we study generalized seeds that contain both, i.e.
    \begin{align} 
     c\zero\hd(1,2,3,4,5) &= 
     f(1,2,3,4,5) + {\rm cyclic},
    \end{align} 
where $f(...)$ is a function of Mandelstam invariants and color structures. 
Independent numerator seeds can give rise to redundant adjoint numerators, as discussed in the 4-point case. 
For example, if $g(...)$ is cyclic in its arguments (just like a single trace), then choosing $f(1,2,3,4,5)=g(1,2,3,5,4)$ generates the same adjoint numerator as a linear combination of 
$c\zeroone\hd(1,2,3,4,5)=g(1,2,3,4,5)$ and 
$c\zerotwo\hd(1,2,3,4,5)=g(1,3,5,2,4)$.
This can be established a priori, in analogy with the fact that $\bar c\zero(1,2,3,4,5) =(123 54)+\text{cyclic}$ is redundant with the two pure color seeds of \autoref{5ptc0s}.
We leave the exploration of the independent basis of seeds for future work.
In practice, it is straightforward to identify the independent numerators from the over-complete set built from seeds.

\subsection{BAS+h.d.\ amplitudes}\label{5ptsMinimalGs}

At any multiplicity, the BAS matrix of doubly color-ordered amplitudes that derives from the generalized numerators can be written as (see \autoref{mhdleftright})
    \beq \label{mhdleftright-3}
        {\bf m}_n\hd = 
        {\bf H}_\textsc{r}\hd \cdot 
        {\bf m}_n \cdot 
        {\bf H}_\textsc{l}\hd\,.
    \eeq 
At 4-point, using generalized gauge transformations, we have previously shown that ${\bf H}\hd_{\textsc{l}/\textsc{r}}$ can be taken to be diagonal matrices, and that these capture all solutions to the KLT bootstrap.
At 5-point, the generalized gauge transformations can reduce any single-trace numerator seed to the linear combination 
\beq \label{seed1}
        c\zero\hd(1,2,3,4,5) = g_1(1,2,3,4,5)\,
                            (12345)\,+g_2(1,2,3,4,5)\,
                          (13524),
\eeq
for independent $g_1$ and $g_2$ functions of the Mandelstam invariants that are cyclically symmetric.
Even if an original seed is free of poles, capturing the same amplitude with $c\zero\hd$ may introduce poles. 
For example, the amplitudes constructed from
    \beq 
    \bar c\zero^{\, \text{h.d.}} (1,2,3,4,5) 
              = h(1,2,3,4,5)\,(12354) + \text{cyclic}\,,
    \eeq 
for any function $h$ depending only on the Mandelstam invariants, may equivalently be constructed from
\beq 
    g_1(1,2,3,4,5) =  
    \frac{
    {-h(3,4,5,2,1)-h(5,1,2,4,3)}
        }{
s_{12}s_{34}
\ m_5[12345|12345]
   }+ \text{cyclic}
   \,,
    \eeq 
and
    \beq 
    g_2(1,2,3,4,5) =  
    \frac{
    {h(3,5,2,1,4)}
        }{
s_{12}s_{34}
\ m_5[12345|12345]
   }+ \text{cyclic}
   \,,
    \eeq
where $m_5[12345|12345]=1/s_{12}s_{34} + \text{cyclic}$, showing that the functions $g_1$ and $g_2$ are in general likely to have poles.
This means that an amplitude built with  $c\zero\hd$ for those special values of $g_{1,2}$ is equal to an amplitude built with  $\bar c\zero\hd$, because both seeds are related by a generalized gauge transformation (nevertheless, the adjoint numerators obtained via these seeds are linearly independent).  
Therefore any 5-point BAS matrix of doubly color-ordered amplitudes can be written as
    \beq \label{gen5ptleftright}
        {\bf m}_5\hd = 
        \left( {\bf G}\hd_{1,\textsc{r}} + {\bf G}\hd_{2,\textsc{r}}\right)
        \cdot {\bf m}_5
        \cdot
        \left( {\bf G}\hd_{1,\textsc{l} }+  {\bf G}\hd_{2,\textsc{l}}\right)\,,
    \eeq 
where ${\bf G}\hd_{1,2}$ follow from the numerator seed defined by \autoref{seed1} by stripping off the single traces. The
${\bf G}\hd_{1}$ is \emph{diagonal}  and 
${\bf G}\hd_{2}$ is \emph{non-diagonal}, with only one non-zero entry on each row/column, as can be seen in \autoref{appJ5pt}.
The non-diagonal matrix is necessary because the 5-point BAS matrix ${\bf m}_5$ contains zero entries, which may become non-zero when higher-derivative corrections are included.

It is non-trivial to verify whether \autoref{gen5ptleftright} covers all solutions to the KLT bootstrap of \citeword\cite{Chi:2021mio}.
As argued in general in \autoref{s:31}, ${\bf m}_5\hd$ has minimal rank just as ${\bf m}_5$ since ${\bf G}\hd_{1,\textsc{r/l}} + {\bf G}\hd_{2,\textsc{r/l}}$ has full rank.
We have reproduced the solution of the KLT bootstrap for all orders explicitly provided in \cite{Chi:2021mio} and the forms of the necessary functions are listed in \autoref{app:solKLT5pt}.
Going beyond this, we also checked that the numerator seeds reproduce the lowest-order 5-point contact terms, which are cubic in the Mandelstam invariants.
These are captured by \autoref{gen5ptleftright} with
    \begin{align}
        g_{1,\textsc{l}}(1,2,3,4,5) &= 1+\frac{c_1}{\Lambda^{10}}\,s^5_{\rm cyclic}\,,\quad 
        g_{2,\textsc{l}}(1,2,3,4,5) = \frac{c_2}{{\Lambda^{10}}}\,s^5_{\rm cyclic}\,,
        \nn\\
        g_{1,\textsc{r}}(1,2,3,4,5) &= 1+\frac{c_3}{\Lambda^{10}}\,s^5_{\rm cyclic}\,,\quad
         g_{2,\textsc{r}}(1,2,3,4,5) = \frac{c_4}{\Lambda^{10}}\,s^5_{\rm cyclic}\,,
    \end{align}
where $s^5_{\rm cyclic}\equiv s_{12}s_{23}s_{34}s_{45}s_{51}$ and $c_i$ are free parameters.
Recall that ${\bf m}_5$ is $\mathcal{O}(1/s_{ab}^2)$ so that the resulting ${\bf m}_5\hd$ is of third order in the Mandelstam invariants.
This exposes a simple structure of higher-order corrections.

\subsection{Factorization properties and particle spectrum}

As discussed in \autoref{s:31}, given a single-copy amplitude $A\gen$ that satisfies the generalized KKBCJ relation, the object $A= ({\bf H}\hd)^{-1}\cdot A\gen$ satisfies the usual KK and BCJ relations (where we momentarily omitted the subscript $L/R$, for simplicity).
Moreover, the same double copy can be constructed using either of these two single copies.
However, it is not immediately clear that $A$ and $A\gen$ share the same analytic properties.
At 4-point, we showed that this is indeed the case.
At 5-point, achieving a fully general proof seems far more challenging.
Therefore, we start the exploration of this question with simplifying assumptions.

Since one can write ${\bf H}\hd = ({\bf G}_{1}+{\bf G}_{2})$ at 5-point, it follows that $A\gen= \left({\bf G}_1+{\bf G}_2\right) \cdot A$.
Restricting to the two independent $12345$ and $13524$ color orderings, we have that
\begin{equation}
\begin{aligned}
    A'[12345] &= g_1(1,2,3,4,5)A[12345]
                +g_2(1,2,3,4,5) A[13524]\,,\\
    A'[13524] &= g_1(1,3,5,2,4)A[13524]
                +g_2(1,3,5,2,4)A[15432]\,.
\end{aligned} 
\end{equation}
This is a simple choice of BCJ basis since the amplitudes do not share any poles (see \autoref{sec:5pt-kernel-from-seeds} for more details).
The full set of ordered amplitudes can be reconstructed using the KKBCJ relations.
The fact that $A[15432]$ satisfies the traditional KK reflection relation $A[15432]=-A[12345]$ allows to write
\begin{align} 
\pmx{A[12345]\\A[13524]} &= 
\pmx{~~g_1(1,2,3,4,5) &  g_2(1,2,3,4,5)\\
     -  g_2(1,3,5,2,4) & g_1(1,3,5,2,4)}^{-1}
     \cdot ~
\pmx{A\gen[12345]\\A\gen[13524]}\,.
\label{eq:curly-g-matrix}
\end{align}
Therefore, the unprimed amplitude can be written as
\beq 
A[12345] = \frac{A\gen[12345]g_1(1,3,5,2,4) + A\gen[13524]g_2(1,3,5,2,4)}
{g_1(1,2,3,4,5)g_1(1,3,5,2,4)+g_2(1,2,3,4,5)g_2(1,3,5,2,4)}\,.
\eeq 
Studying the analytic properties of the amplitude $A$ is challenging for two main reasons.
First, as discussed in \autoref{5ptsMinimalGs} and in contrast to the 4-point case, the functions $g_1$ and $g_2$ may contain poles.
So a general parametrization has a complicated form and studying whether the poles of $A$ and $A\gen$ agree is non-trivial.
Second, even if we take analytic $g_1,\,g_2$ so that $A$ and $A\gen$ poles are identical, a suitable redefinition of $A$ may be necessary to guarantee that its residues match those of $A\gen$ (similarly to the 4-point case presented in \autoref{gentonongenR}).  

Working out such a redefinition at 5-point, or even proving that one always exists, is beyond the scope of this paper.
However, in the simplest setup, namely that of a kernel whose first EFT correction is a 5-point contact term constructed from a vanishing $g_2$ and a polynomial $g_1$, we can identify a suitable redefinition of $A_\textsc{l/r}$ such that their residues agree with those of $A\gen_\textsc{l/r}$ while leaving the double copy unchanged (we checked this property up to order $\cO(1/\Lambda^{10})$).
Explicitly, we found that the locality properties of the kernel at 5-point impose that (for seeds up to order $\cO(1/\Lambda^{10})$)
\beq\label{diagonalPolys5Points}
g_{1,\textsc{l/r}}(1,2,3,4,5)=p_\textsc{l/r}(1,2,3,4,5)+\frac{c_\textsc{l/r}}{\Lambda^{10}}s_{12}s_{23}s_{34}s_{45}s_{51} \ ,
\eeq
where $c_\textsc{l/r}$ are free constants and $p_\textsc{l/r}(1,2,3,4,5)=p_\textsc{l/r}$ are permutation-invariant functions such that $p_\textsc{l}p_\textsc{r}=1$ whenever a Mandelstam invariant vanishes.
Consequently,
\beq
p_\textsc{l}A_\textsc{l}[\alpha]
\quad\text{ and }\quad
\frac{A_\textsc{r}[\alpha]}{p_\textsc{l}}\,,
\eeq
have the same residues on the poles as $A\gen_\textsc{l}$ and $A\gen_\textsc{r}$, respectively.
Therefore, the double-copy amplitudes associated to the kernel obtained from the seeds of \autoref{diagonalPolys5Points} can also be obtained with the traditional KLT kernel and single-copy amplitudes which verify the usual KKBCJ relations.
Whether such manipulations can be performed in full generality at 5-point is a question left for future investigation.

\section{Conclusions and outlook}

In this paper, we revisited recent proposals for the systematic double copy of effective field theories.
Inspired by the decomposition of adjoint color factors into single traces of Lie algebra generators, we proposed a method to construct generalized adjoint numerators.
These satisfy Jacobi-like and antisymmetry relations, while depending on both color and kinematics.
Starting from \emph{numerator seeds} satisfying the permutation properties of single traces, we proved that all adjoint numerators can be obtained through the linear map~${\bf J}$ between single-trace and adjoint color factors.
While generalized numerators have previously been constructed up to 5-point in \cite{Carrasco:2019yyn, Carrasco:2021ptp}, the construction from numerator seeds is advantageous because the algebraic properties of single traces are simpler than the adjoint algebraic relations. 
We showed that this method works for any multiplicity and it is convenient to explore the higher-derivative corrections that allow for a color-dual representation.

The matrix ${\bf J}$ is also instrumental in relating the amplitudes represented with trivalent graphs involving (generalized) adjoint numerators to color-ordered ones.
The construction of generalized adjoint numerators therefore facilitates the comparison between the generalized numerators construction of \cite{Carrasco:2019yyn, Carrasco:2021ptp} and the generalized KLT formalism of \cite{Chi:2021mio}. 
At 4-point, we showed that the generalized adjoint numerators encode all the KLT bootstrap solutions, to any order in the EFT expansion. 
The two approaches therefore allow for exactly the same higher-derivative corrections to the bi-adjoint scalar amplitudes.
The single-copy amplitudes are also the same in the two formalisms.
This insight consequently exposed the structure of double-copy amplitudes.
While the generalized KLT formalism does expand the range of operators in the single-copy amplitudes, we find (at 4-point) that any resulting double copy can also be obtained with the traditional KLT kernel.
We provide partial 5-point results suggesting that these conclusions may extend to higher multiplicity.
However, due to the more complicated KLT bootstrap and structures involved, further investigations on higher multiplicities are left to future work.

There are several directions that deserve further attention.
For example, we have focused on the construction of scalar numerators, as opposed to gauge-theory numerators involving polarization vectors. 
The relevant purely kinematic numerator seeds have been considered for Yang--Mills theory in~\cite{Bern:2011ia, Bjerrum-Bohr:2012kaa, Du:2013sha, Fu:2013qna,Naculich:2014rta}, but not in the EFT context or for generalized numerators. 
Gauge invariance must hold at the level of amplitudes (i.e.~$\mathcal{A}|_{\eps_i\to p_i}=0$), and individual entries in a numerator are typically not gauge invariant.
The necessary extra constraint on numerator seeds may therefore take a complicated form, especially beyond 4-point.
Methods proposed to identify the possible gauge-invariant structures in Yang--Mills theories \cite{Boels:2016xhc, Bern:2017tuc} could be useful. 
In particular, a basis of cyclically invariant structures which can be used as numerator seeds was provided in \citeword\cite{Bern:2017tuc}.

Furthermore, we have only considered color factors consisting of single traces of Lie algebra generators. 
However, the construction of adjoint numerators from numerator seeds does apply more generally. For instance, at 4-point there exist theories with double traces, which can be combined into the seed $c\zero(1,3,2,4) =\Tr(T^{a_1}T^{a_2})\Tr(T^{a_3}T^{a_4}) = \delta^{12}\delta^{34}$. 
The resulting generalized adjoint numerator was previously identified in \citeword\cite{Low:2020ubn}.
Constructing generalized color factors involving products of traces may lead to interesting new single or double copies, which would also be worth studying in the generalized KLT formalism.

Besides the assumption of single traces, we have not included the possibility of extra particles beyond the bi-adjoint scalar discussed in \autoref{s:doublecopystructure}. 
For the double copy of a gauge theory with matter in the fundamental representation, the KLT kernel is for instance constructed from a \textit{bicolor} theory containing two scalars \cite{delaCruz:2016wbr, Brown:2018wss}.
Including new particles in factorization channels, or even externally, in the generalized KLT formalism would allow for a larger space of single-copy amplitudes.
It would then be worthwhile to extend the numerator seeds to amplitudes with a more complicated particle spectrum.

Altogether, the simple construction of adjoint numerators from numerator seeds has been useful to explore the structures in the generalized CK and KLT double-copy formalisms. 
Still, the double copy of effective field theories retains various unexplored aspects which promise exciting new findings for the years to come.

\section*{Acknowledgments}

We thank Jan Plefka and Johannes Broedel for helpful discussions. This work is supported by the Deutsche Forschungsgemeinschaft under Germany's Excellence Strategy  EXC 2121 ``Quantum Universe'' - 390833306.
J.R.N.\ is supported by the Deutsche Forschungsgemeinschaft (DFG, German Research Foundation) - Projektnummer 417533893/GRK2575 “Rethinking Quantum Field Theory”.

\appendix

\section{Proof: numerator seeds construct all adjoint numerators} \label{proof}
In this appendix, we prove that any adjoint numerator can be constructed from a numerator seed, as in \autoref{seedeq}.
For any (not necessarily square) matrix ${\bf J}$, there exists a unique \textit{pseudoinverse}
${\bf J}^{+}$ (see for instance \cite{Ben-Israel1980}) satisfying
\beq \label{pseudoinv}
{\bf J}\cdot {\bf J}^{+}\cdot {\bf J} = {\bf J}
\eeq
(together with 
${\bf J}^{+}\cdot {\bf J}\cdot {\bf J}^{+} = {\bf J}^{+},$
$({\bf J}\cdot {\bf J}^{+})^\transpose = {\bf J}\cdot {\bf J}^{+},$ and
$({\bf J}^{+}\cdot {\bf J})^\transpose = {\bf J}^{+}\cdot {\bf J}$ for real ${\bf J}$, which will be of less importance in the following)
That is, ${\bf J}\cdot {\bf J}^{+}$ maps all columns of ${\bf J}$ to themselves.
Another word for pseudoinverse is the Moore--Penrose inverse. It can be obtained, for example, by the rank decomposition as in \autoref{eq:rankdecomp}.
Here $(2n-5)!!$ is the number of trivalent graphs.  
We then have 
\beq 
    {\bf J}^+ = 
    {\bf B}^\transpose\cdot({\bf B}\cdot{\bf B}^\transpose)^{-1}
    \cdot
    ({\bf A}^\transpose\cdot{\bf A})^{-1}\cdot{\bf A}^\transpose\,.
\eeq 
At 4- and 5-point, the pseudoinverse takes a particularly simple form:
${\bf J}_4^{+}={\bf J}_4^\transpose/6$
and ${\bf J}_5^{+}={\bf J}_5^\transpose/20$.
At 6-point, the pseudoinverse is not proportional to ${\bf J}_6^\transpose$, but can still be obtained algorithmically.

Since the adjoint color factor $c_\text{adj}$ is constructed by ${\bf J}$, \autoref{pseudoinv} allows to express the Jacobi identities in matrix form,
    \beq
       (\mathbbm{1}-{\bf J}\cdot {\bf J}^+)\cdot\vec{c}_{\text{adj}} = 0\,.
    \eeq
By definition, (generalized) color-kinematics duality states that adjoint numerators should obey the same algebraic relations as $c_\text{adj}$. 
In other words, as pointed out in e.g.\ \cite{Momeni:2020hmc,Naculich:2014rta},
 it implies that all adjoint numerators should live in the null-space of $\mathbbm{1}-{\bf J}\cdot {\bf J}^+$. 
Therefore, \textit{any} adjoint numerator 
$\vec{n}$ (with only kinematics or both color and kinematics) that satisfies the Jacobi relations obeys
	\begin{equation}\label{eq:projectnum}
		{\bf J}\cdot {\bf J}^{+} \cdot \vec{n}= \vec{n}\,.
	\end{equation}
This crucially implies that any adjoint numerator at any multiplicity can be written as the ${\bf J}$-matrix multiplying a vector, 
\beq \label{eq:zerothnumnpoint}
    \vec{n}\zero = {\bf J}^{+}\cdot \vec{n}\,.
\eeq
Below, we show that $\vec{n}\zero$ constructed in this way transforms in the same way as $c\zero$ under permutations of the particle labels. It is thus a numerator seed.
Therefore, all adjoint numerators can be constructed from numerator seeds by multiplication with ${\bf J}$.

\paragraph{\texorpdfstring{$ \vec{n}\zero = {\bf J}^{+}\cdot \vec{n}$}{n0 = J+ . n} is a numerator seed.}
\label{seedproof}

In this subsection, we prove that the numerator seeds constructed through \autoref{eq:zerothnumnpoint}, transform according to the same rule as the color factor $c\zero$, 
\beq
c\zero\xrightarrow[\sigma]{} 
M_{c\zero,\sigma} \cdot c\zero \,,
\eeq
under a permutation of the particle labels, $\sigma$. 

First, note that adjoint numerators transform analogously under $\sigma$,
    \beq 
        n_\text{adj} \xrightarrow[\sigma]{} M_{c_\text{adj},\sigma} \cdot n_\text{adj} \,. 
    \eeq 
with the same $M_{c_\text{adj},\sigma} \left(\neq M_{c\zero,\sigma} \right)$ for any adjoint numerator. 
Since the entries of $c\zero$ are linearly independent, 
    \begin{align} 
        c_\text{adj} = {\bf J}\cdot c\zero 
        \ \xrightarrow[\sigma]{} \ 
       M_{c_\text{adj},\sigma} \cdot {\bf J}\cdot c\zero  
        &= {\bf J}\cdot M_{c\zero,\sigma} \cdot c\zero 
\nn\\\implies
         M_{c_\text{adj},\sigma} \cdot {\bf J}
         &= {\bf J}\cdot M_{c\zero,\sigma} \,.
    \end{align} 
The numerator seeds transform as
\beq 
    {n}\zero = {\bf J}^{+}\cdot {n}_\text{adj}
    \ \xrightarrow[\sigma]{} \ 
    {\bf J}^{+} \cdot 
        M_{c_\text{adj},\sigma} \cdot n_\text{adj} 
    \,,
\eeq
which we want to show is the same as 
$   {\bf J}^{+}\cdot {n}_\text{adj}
    \ \xrightarrow[\sigma]{?} \ 
    M_{c\zero,\sigma} \cdot {\bf J}^{+} \cdot 
          n_\text{adj} 
$.
We will prove this by showing that 
$\left[{\bf J}^{+}{\bf J}, M_{c\zero,\sigma} \right] = 0$.

Each row and each column of the matrix $M_{c\zero,\sigma}$ contains exactly one $1$ and the rest $0$'s. Such matrices are orthogonal, $M^\transpose_{c\zero,\sigma} = M^{-1}_{c\zero,\sigma}$.
In addition, any permutation can be written as a product of transpositions (the interchange of two labels, $\sigma=a\leftrightarrow b$). Therefore, it will be enough to prove that 
$\left[{\bf J}^{+}{\bf J}, M_{c\zero,a\leftrightarrow b} \right] = 0$.
Performing the same transposition twice is the same as doing nothing. That is, transpositions are their own inverse, 
    \beq \label{Mproperties}
        M_{c\zero,a\leftrightarrow b} = 
        M^{-1}_{c\zero,a\leftrightarrow b} = 
        M^\transpose_{c\zero,a\leftrightarrow b}\,.
    \eeq

As in \autoref{eq:rankdecomp}, ${\bf J}$ can be decomposed as ${\bf J}={\bf A}\cdot{\bf B}$, where ${\bf B}$ is the matrix that relates $c\zero$ to some choice of DDM basis, $c_\textsc{ddm} = {\bf B}\cdot c\zero$. 
This implies
    \beq 
        {\bf J}^+\cdot {\bf J} = 
        {\bf B}^\transpose\cdot
        ({\bf B}\cdot{\bf B}^\transpose)^{-1}
        \cdot {\bf B}
        \,.
    \eeq
The entries in the DDM basis correspond to all half-ladder graphs with fixed endpoints. 
If $a,b$ are not those endpoints, then 
$\sigma=a \leftrightarrow b$ maps entries of the DDM basis to other entries of the same DDM basis, so
$c_\textsc{ddm} \xrightarrow[\sigma]{} 
    M_{c_\textsc{ddm},a \leftrightarrow b} \cdot c_\textsc{ddm}$.
Therefore for each transposition there exists a choice of ${\bf B}$ such that 
    \begin{align}
        {\bf B}\cdot c\zero 
        \xrightarrow[\sigma]{} 
    M_{c_\textsc{ddm},a \leftrightarrow b} \cdot {\bf B}\cdot c\zero &=
    {\bf B}\cdot M_{c\zero,a \leftrightarrow b}  
    \cdot c\zero \nn\\\label{MBcommute}
\implies
      M_{c_\textsc{ddm},a \leftrightarrow b} \cdot {\bf B}
     &= 
       {\bf B}\cdot M_{c\zero,a \leftrightarrow b} \,.
    \end{align}

Now, using \autoref{Mproperties} and \autoref{MBcommute}, 
it is straightforward to show that 
    \beq 
        M_{c\zero,a \leftrightarrow b}
        \cdot {\bf B}^\transpose\cdot
        ({\bf B}\cdot{\bf B}^\transpose)^{-1}
        \cdot {\bf B}
        =
                {\bf B}^\transpose\cdot
        ({\bf B}\cdot{\bf B}^\transpose)^{-1}
        \cdot {\bf B} \cdot M_{c\zero,a \leftrightarrow b} 
        \,.
    \eeq 
Therefore 
    \begin{align} 
      M_{c\zero,\sigma} \cdot {\bf J}^{+}\cdot {\bf J}
    =  {\bf J}^{+}\cdot &{\bf J}\cdot M_{c\zero,\sigma}
    = {\bf J}^{+} \cdot M_{c_\text{adj},\sigma}\cdot{\bf J} 
    \nn\\[2mm]
    \implies 
    {\bf J}^{+} \cdot M_{c_\text{adj},\sigma} \cdot
    n_\text{adj} 
    &= 
    {\bf J}^{+} \cdot M_{c_\text{adj},\sigma}\cdot{\bf J} \cdot{\bf J}^+ \cdot 
    n_\text{adj}
    \nn\\
    &= 
   M_{c\zero,\sigma}\cdot {\bf J}^{+} \cdot {\bf J} \cdot{\bf J}^+ \cdot 
    n_\text{adj}
    \nn\\
    &= M_{c\zero,\sigma}\cdot {\bf J}^{+} \cdot 
    n_\text{adj} \,.
    \end{align}
Thus, as we wanted to show, the numerator seed
$n\zero = {\bf J}^{+}\cdot {n}_\text{adj}$ (which exists for any adjoint numerator) transforms as
\beq 
n\zero
    \ \xrightarrow[\sigma]{} \ 
    M_{c\zero,\sigma} 
    \cdot n\zero \,.
\eeq

\section{Examples of seeds and adjoint numerators }\label{exnums}
In this appendix, we present the construction of kinematic adjoint numerators from numerator seeds. 
In general, a numerator seed can be obtained from any function $f(1,2,...,n)$ through
\beq 
    n\zero(1,2,...,n) = 
    \big( f(1,2,...,n) + (-1)^n\,f(n,...,2,1) \big)
    + \text{cyclic}\,.
\label{decompositionSeeds}    
\eeq
The number of independent numerator seeds built using this equation, up to 6-point and dimension 30, is provided in \autoref{table-seed-counting}.
After multiplying them by ${\bf J}$, one needs to explicitly verify the linear independence of the resulting adjoint numerators. 
To determine the counting provided in \autoref{table6pt}, we first construct vectors by evaluating the numerators numerically for different values of the momenta. 
The rank of the matrix formed with these vectors as columns is the number of independent adjoint numerators.

A systematic correspondence can be established between the entries of \autoref{table6pt} and \autoref{table-seed-counting}.
At 5-point, the number of independent numerators is, for instance, exactly half of that of independent seeds.
This can be understood from the algebraic properties of color factors decomposed into structure constants $f^{abc}$ and symmetric $d^{abc...}$, as discussed in~\cite{Carrasco:2021ptp} up to 5-point.
The 4- and 5-point cases are detailed below.

\begin{table}[t]
    \centering
    \begin{tabular*}{\textwidth}{@{\extracolsep{\fill}}|c||*{15}{@{\:}c@{\;}|}}
         \hline
          {\bf k} 
          &\bf 1&\bf 2& \bf 3& \bf 4&\bf 5&\bf 6&\bf 7&\bf 8&\bf 9&
          \bf 10&\bf 11&\bf 12&\bf 13&\bf 14&\bf 15 
          \\\hline\hline
         {\bf 4-pt} 
         &1&2&2&3&3&4&4&5&5&6&6&7&7&8&8
         \\\hline
         {\bf 5-pt} 
         &0&0&2&4&10&16&28&42&64&90&126&168&224&288&370
         \\\hline
         {\bf 6-pt} 
         &2&8&22&58&133&298&600&1166
         &2132&3754&6324&10351&16368&
         25266&38004
         \\\hline
    \end{tabular*}
    \caption{%
    Counting of scalar kinematic numerator seeds up to 6-point and at $\mathcal{O}(1/\Lambda^{2k})$ for $k\le15$ in the EFT expansion.
    The Gram determinant constraints relevant in 4 spacetime dimensions have been accounted for.
   }
    \label{table-seed-counting}
\end{table}

\paragraph{4-point.}
At 4-point, the numerator seeds are given by \autoref{expansiong}. At lowest orders, we have
\begin{equation}
\begin{aligned}
    n\zero(1,3,2,4) = s 
&\implies 
    n_{\text{adj},s} = t-u\,,
    \\
    n\zero(1,3,2,4) = s^2 
&\implies 
    n_{\text{adj},s} = t^2-u^2 = s(u-t)\,.
\end{aligned}
\end{equation}
The other valid numerator seed at second order, 
$n\zero(1,2,3,4)=t\,u$ maps to the same adjoint numerator, because the permutation invariant $s^2+t^2+u^2=2(s^2-t\,u)$ maps to zero. At third order (and any higher order), the adjoint numerators are permutation invariant functions multiplying the lowest two orders \cite{Carrasco:2019yyn}:
\begin{align}
    n\zero(1,3,2,4) = s^3 
&\implies 
    n_{\text{adj},s} = t^3-u^3 = (t-u)(s^2+t^2+u^2)/2\,.
\end{align}
Upon comparing \autoref{table6pt} and \autoref{table-seed-counting}, one can identify the difference between the numbers of independent seeds and numerators at 4-point as being the number of permutation-invariant functions of the Mandelstam invariants.
The latter are combinations of terms of the form $(s^2+t^2+u^2)^m(stu)^n$ for integer $m,n$.
Indeed, the seeds at 4-point are symmetric under reversal symmetry, hence they capture all expressions which are combinations of reversal-symmetric terms of the form
\beq
f(\sigma(a),\sigma(b),\sigma(c),\sigma(d))+f(\sigma(d),\sigma(c),\sigma(b),\sigma(a))
\eeq
for some permutation $\sigma$ and some function $f$, consistently with \autoref{decompositionSeeds}.
They do however not capture expressions that are antisymmetric under the reversal of particle labels.
To understand what algebraic structures appear in the symmetric combinations, one can consider the permutation properties of color structures generated by single traces of group generators, which are identical to those of $f(1,2,...,n)+\text{cyclic}$ in \autoref{decompositionSeeds}.
At 4-point, the relevant color structures are $d^{abx}f^{xcd},f^{abx}f^{xcd}$ and $d^{abcd}$~\cite{Carrasco:2021ptp}.
The first one is expressed in terms of antisymmetric combinations of traces only, whereas the other two are expressed in terms of symmetric combinations.
Therefore, the seeds generate expressions with the same algebraic properties as the two last color structures, which are respectively adjoint and permutation-invariant.

\paragraph{5-point.}
At the lowest two orders in the Mandelstam invariants, there are no numerator seeds.
For example, the candidate function $f(1,2,3,4,5) = s_{12}+s_{23}+s_{34}+s_{45}+s_{51}$ is invariant under cyclic permutations, but symmetric instead of antisymmetric under reversing the order of its arguments, therefore it is mapped to zero under the action of $\bf J$.  
At third order in the Mandelstams, there exists
\beq 
    n\zero(1,2,3,4,5) = \frac{1}{2}(s_{12}^2\,s_{34}-s_{45}^2\,s_{23})
    +\text{cyclic}
                        \,,
\eeq 
which is mapped to the adjoint numerator
\begin{align} 
    n_{\text{adj},s_{12}s_{45}} = 
 \hspace{3mm}&
 \big(n\zero(1,2,3,4,5)
  -n\zero(1,5,4,3,2)\big)
 +\big(n\zero(1,4,5,3,2)
 -n\zero(1,2,3,5,4)\big) \nn\\
 +&\big(n\zero(1,2,5,4,3)
 -n\zero(1,3,4,5,2)\big)
 +\big(n\zero(1,3,5,4,2)
 -n\zero(1,2,4,5,3)\big)
\nn \\[1.5mm]
=\hspace{3mm}&(s_{24} s_{35}^2
-s_{24} s_{13}^2)+
\big(s_{24}^2 s_{13}
-s_{24}^2 s_{35}\big)+
\big(s_{25} s_{13}^2
-s_{14} s_{35}^2\big) \nn\\+&
\big(s_{14}^2 s_{35}
-s_{25}^2 s_{13}\big)+
\big(s_{14} s_{23}^2
-s_{25} s_{34}^2\big)+
\big(s_{25}^2 s_{34}
-s_{14}^2 s_{23}\big)\nn\\+&
\big(s_{15}s_{34}^2
-s_{15} s_{23}^2\big)+
\big(s_{15}^2 s_{23}
-s_{15}^2 s_{34}\big)\,.
\label{example5pt3rdorder}
\end{align}
Here, the subscript $s_{12}s_{45}$ indicates the pole structure of the trivalent graph to which this entry belongs, i.e.\ 
$c_{\text{adj},s_{12}s_{45}} = f^{12x}f^{x3y}f^{y45}$.
The first line in \autoref{example5pt3rdorder} is analogous to
\begin{equation}
\begin{aligned} 
 f^{12x}f^{x3y}f^{y45} = 
 \hspace{3mm}&
 \big((12345)
  -(15432)\big)
 +\big((14532)
 -(12354)\big)\\
 +&\big((12543)
 -(13452)\big)
 +\big((13542)
 -(12453)\big)\,.
\end{aligned}
\end{equation}
(recall $(12...5)\equiv\text{Tr}(T^{a_1}T^{a_2}...T^{a_5})$, etc.).

The comparison between \autoref{table6pt} and \autoref{table-seed-counting} shows that the number of independent adjoint numerators is exactly half that of independent seeds.
As in the 4-point case, this can be understood from the algebraic properties of the color factors which are generated by combinations of single traces of group generators with the same behavior under reversal symmetry as the seeds.
At 5-point, the seeds are antisymmetric under reversal, while the color factors which can be decomposed onto antisymmetric combinations of single traces can be identified from the classification in~\cite{Carrasco:2021ptp}.
In the language of this reference, they correspond to adjoint and hybrid structures, which correspond to combinations of color factors of the form $f^{abx}f^{xcy}f^{yde}$ and $d^{abcx}f^{xde}$.
Therefore, antisymmetric seeds generate all expressions having the algebraic properties of these two kinds.
In addition, there exists a bijection between adjoint and hybrid structures~\cite{Carrasco:2021ptp}.
Namely, for each expression with adjoint properties, there exists one with hybrid properties, and reciprocally.
It then follows that the number of independent seeds is exactly twice that of independent adjoint numerators.

\paragraph{6-point.}
At lowest order, there exist two functional forms that are cyclically invariant and symmetric under argument reversal,
\begin{equation}
\begin{aligned}
    n\zero(1,2,3,4,5,6) &= 
    s_{12}+s_{23}+s_{34}+s_{45}+s_{56}+s_{16}\,,\\
    n\zero(1,2,3,4,5,6)&= 
    s_{123}+s_{234}+s_{345}
    \,,
\end{aligned}
\end{equation}
where $s_{abc}=(p_a+p_b+p_c)^2$.
These map to one independent adjoint numerator after multiplication by ${\bf J}$.
At second order, there are three independent adjoint numerators.
Out of the eight independent seeds, these can for instance be constructed from
    \begin{equation}
        \begin{aligned}
    n\zero(1,2,3,4,5,6) &= 
    s_{123}\,(s_{234}+s_{345})
    +s_{234}\,s_{345}\,, \\
    n\zero(1,2,3,4,5,6) &= 
    s_{16} s_{34}+
    s_{12} s_{45}+
    s_{23} s_{56}\,, \\
    n\zero(1,2,3,4,5,6) &= 
    s_{123}(s_{16}+s_{34})+
    s_{345}(s_{23}+s_{56})+
    s_{234}(s_{12}+s_{45}) \,. 
        \end{aligned}
    \end{equation} 
The adjoint numerator then has the form
    \begin{align}
        n_{\text{adj},s_{12}s_{34}s_{56}}
        &=
        \big[ n\zero(1, 2, 3, 4, 5, 6) 
        - n\zero(1, 2, 3, 4, 6, 5)
        -  n\zero(1, 2, 4, 3, 5, 6) 
        + n\zero(1, 2, 4, 3, 6, 5)\nn \\
        &\;\;-  n\zero(1, 2, 5, 6, 3, 4) 
        + n\zero(1, 2, 5, 6, 4, 3) + 
 n\zero(1, 2, 6, 5, 3, 4)
 - n\zero(1, 2, 6, 5, 4, 3)\big]\nn\\
 &\;\;+ \text{reversed orderings}
        \,,
    \end{align}
for the adjoint entry corresponding to
$c_{\text{adj},s_{12}s_{34}s_{56}} = f^{12x}f^{34y}f^{56z}f^{xyz}$, and
    \begin{align}
        n_{\text{adj},s_{12}s_{123}s_{56}}
        &=
        \big[ 
        n\zero(1,2,3,4,5,6)-n\zero(1,2,3,4,6,5)
        -n\zero(1,2,3,5,6,4)+n\zero(1,2,3,6,5,4)\nn \\
        &\;\;-n\zero(1,2,4,5,6,3)+n\zero(1,2,4,6,5,3)
        +n\zero(1,2,5,6,4,3)-n\zero(1,2,6,5,4,3)
        \big]\nn\\
 &\;\;+ \text{reversed orderings}
        \,,
    \end{align}
for the adjoint entry corresponding to
$c_{\text{adj},s_{12}s_{123}s_{56}} = f^{12x}f^{x3y}f^{y4z}f^{z56}$.
At third order, there are nine independent adjoint numerators. They can be constructed from, for example,
\begin{align}
    n\zero(1,2,3,4,5,6) &= 
s_{1  2} s_{3  4} s_{5  6}+ s_{2  3} s_{4  5}s_{1  6}\,,\\
n\zero(1,2,3,4,5,6) &= 
s_{1  2  3} s_{2  3  4} s_{3  4  5}\,,\nn\\
n\zero(1,2,3,4,5,6) &= 
s_{1  2}s_{2  3} s_{3  4} 
+s_{1  2}s_{2  3} s_{1  6} 
+s_{2  3} s_{3  4} s_{4  5} 
+s_{1  2} s_{1  6} s_{5  6}
+s_{1  6} s_{4  5}
   s_{5  6}+s_{3  4} s_{4  5} s_{5  6}\,,\nn\\
n\zero(1,2,3,4,5,6) &= 
(s_{5  6} s_{1  2}^2
+s_{3  4}^2 s_{1  2}+s_{1  6} s_{2  3}^2+s_{2  3} s_{4  5}^2+s_{3  4} s_{5  6}^2+s_{1  6}^2 s_{4  5})\nn\\
&\quad +(1\leftrightarrow6,~2\leftrightarrow5,~3\leftrightarrow4)\,,\nn\\
n\zero(1,2,3,4,5,6) &= 
s_{234} (s_{12}^2+s_{45}^2)+
 s_{123}(s_{16}^2+s_{34}^2)+
 s_{345}(s_{23}^2+s_{56}^2)
\,,\nn\\
n\zero(1,2,3,4,5,6) &= 
 s_{123} s_{345}(s_{12}+s_{45} )+
 s_{123} s_{234}(s_{23}+s_{56})+
s_{234}s_{345} (s_{16}+s_{34}) 
\,,\nn\\
n\zero(1,2,3,4,5,6) &= 
s_{123}^2 \left(s_{234}+s_{345}\right)+
s_{234}^2 \left(s_{123}+s_{345}\right)+
s_{345}^2 \left(s_{123}+s_{234}\right)
\,,\nn\\
n\zero(1,2,3,4,5,6) &= 
\left(s_{123}+s_{234}+s_{345}\right){}^3
\,,\nn\\
n\zero(1,2,3,4,5,6) &= 
\left(s_{12}+s_{23}+s_{34}+s_{45}+s_{56}+s_{16}\right){}^3
\,.\nn
\end{align}

We leave the detailed comparison between the entries of \autoref{table6pt} and \autoref{table-seed-counting} at 6-point (and higher) to future work.

\section{Details at 5-point}

\subsection{Matrices}\label{appJ5pt}

At 5-point, for the trace ordering given by
\begin{equation}
\begin{aligned}
    c\zero = \big(
    \, &(12345),
    (12354),
    (12435),
    (12453),
    (12534),
    (12543),
    (13245),
   (13254),\\
   &(13425),
   (13524),
   (14235),
   (14325),
   (15432),
   (14532),
   (15342),
   (13542),\\
   &(14352),
   (13452),
   (15423),
   (14523),
   (15243),
   (14253),
   (15324),
   (15234) \big)^\transpose\,,
\end{aligned}
\end{equation}
and the adjoint color numerators by
\begin{equation}
\begin{aligned}
    c_\text{adj}=
    \Big( \ &f^{12x} f^{34y} f^{5xy},
    f^{12x} f^{35y} f^{4xy},
    f^{12x} f^{45y}
   f^{3xy},
   f^{13x} f^{24y} f^{5xy},\\& 
   f^{13x} f^{25y} f^{4xy},
   f^{13x}
   f^{45y} f^{2xy},
   f^{14x} f^{23y} f^{5xy},
   f^{14x} f^{25y}
   f^{3xy},\\&
   f^{14x} f^{35y} f^{2xy},
   f^{15x} f^{23y} f^{4xy},
   f^{15x}
   f^{24y} f^{3xy},
   f^{15x} f^{34y} f^{2xy},\\& f^{23x} f^{45y}
   f^{1xy},f^{24x} f^{35y} f^{1xy},f^{25x} f^{34y} f^{1xy} \ 
   \Big)^\transpose,
\end{aligned}
\end{equation}
the matrix ${\bf J}_5=\Big( {\bf J}_{15\times 12}, -{\bf J}_{15\times 12}\Big)^\transpose$ with
\beq 
\scalebox{.82}{$
{\bf J}_{15\times 12} = 
\left(
\begin{array}{cccccccccccc}
 1 & 0 & -1 & 0 & -1 & 1 & 0 & 0 & 0 & 0 & 0 & 0 \\
 0 & 1 & -1 & 1 & -1 & 0 & 0 & 0 & 0 & 0 & 0 & 0 \\
 -1 & 1 & 0 & 1 & 0 & -1 & 0 & 0 & 0 & 0 & 0 & 0 \\
 0 & 0 & 0 & -1 & 0 & 0 & 1 & 0 & -1 & -1 & 0 & 0 \\
 0 & 0 & 0 & 0 & 0 & -1 & 0 & 1 & -1 & -1 & 0 & 0 \\
 0 & 0 & 0 & 1 & 0 & -1 & -1 & 1 & 0 & 0 & 0 & 0 \\
 0 & -1 & 0 & 0 & 0 & 0 & 0 & 1 & 0 & 0 & 1 & -1 \\
 0 & 0 & 0 & 0 & -1 & 0 & 0 & 1 & 0 & -1 & 0 & -1 \\
 0 & 1 & 0 & 0 & -1 & 0 & 0 & 0 & 0 & -1 & -1 & 0 \\
 -1 & 0 & 0 & 0 & 0 & 0 & 1 & 0 & 0 & 0 & 1 & -1 \\
 0 & 0 & -1 & 0 & 0 & 0 & 1 & 0 & -1 & 0 & 1 & 0 \\
 1 & 0 & -1 & 0 & 0 & 0 & 0 & 0 & -1 & 0 & 0 & 1 \\
 1 & -1 & 0 & 0 & 0 & 0 & -1 & 1 & 0 & 0 & 0 & 0 \\
 0 & 0 & 1 & -1 & 0 & 0 & 0 & 0 & 0 & -1 & -1 & 0 \\
 0 & 0 & 0 & 0 & 1 & -1 & 0 & 0 & -1 & 0 & 0 & 1 \\
\end{array}
\right).$}
\eeq 
A block structure in ${\bf J}_5$ has been made manifest by ordering $c\zero$ into the schematic form: ${c\zero} = \left({c\zero}_{12}, \text{reversed}({c\zero}_{12})\right)^\transpose$, where ${c\zero}_{12}$ contains 12 entries of $c\zero$ that are not related by reversing the order of the particle labels.

The matrix $ {\bf G}_2$, which can be obtained from the numerator seed \autoref{seed1} by stripping of the single traces, for $\tilde{g}_{12345}\equiv g_2(1,2,3,4,5)$, is
\begin{equation}
\adjustbox{max width=\textwidth}{$
{\bf G}_2=
\arraycolsep=-2.3pt
\left(
\begin{array}{cccccccccccccccccccccccc}
 0 & 0 & 0 & 0 & 0 & 0 & 0 & 0 & 0 & \tilde g_{12345} & 0 & 0 & 0 & 0 & 0 & 0 & 0 & 0 & 0 & 0
   & 0 & 0 & 0 & 0 \\
 0 & 0 & 0 & 0 & 0 & 0 & 0 & 0 & \tilde g_{12354} & 0 & 0 & 0 & 0 & 0 & 0 & 0 & 0 & 0 & 0 & 0
   & 0 & 0 & 0 & 0 \\
 0 & 0 & 0 & 0 & 0 & 0 & 0 & 0 & 0 & 0 & 0 & 0 & 0 & 0 & 0 & 0 & 0 & 0 & 0 & \tilde g_{12435}
   & 0 & 0 & 0 & 0 \\
 0 & 0 & 0 & 0 & 0 & 0 & 0 & 0 & 0 & 0 & 0 & \tilde g_{12453} & 0 & 0 & 0 & 0 & 0 & 0 & 0 & 0
   & 0 & 0 & 0 & 0 \\
 0 & 0 & 0 & 0 & 0 & 0 & 0 & 0 & 0 & 0 & 0 & 0 & 0 & 0 & 0 & 0 & 0 & 0 & \tilde g_{12534} & 0
   & 0 & 0 & 0 & 0 \\
 0 & 0 & 0 & 0 & 0 & 0 & 0 & 0 & 0 & 0 & 0 & 0 & 0 & 0 & 0 & 0 & 0 & 0 & 0 & 0 & 0 & 0 &
   \tilde g_{12543} & 0 \\
 0 & 0 & 0 & 0 & \tilde g_{13245} & 0 & 0 & 0 & 0 & 0 & 0 & 0 & 0 & 0 & 0 & 0 & 0 & 0 & 0 & 0
   & 0 & 0 & 0 & 0 \\
 0 & 0 & \tilde g_{13254} & 0 & 0 & 0 & 0 & 0 & 0 & 0 & 0 & 0 & 0 & 0 & 0 & 0 & 0 & 0 & 0 & 0
   & 0 & 0 & 0 & 0 \\
 0 & 0 & 0 & 0 & 0 & 0 & 0 & 0 & 0 & 0 & 0 & 0 & 0 & \tilde g_{13425} & 0 & 0 & 0 & 0 & 0 & 0
   & 0 & 0 & 0 & 0 \\
 0 & 0 & 0 & 0 & 0 & 0 & 0 & 0 & 0 & 0 & 0 & 0 & \tilde g_{13524} & 0 & 0 & 0 & 0 & 0 & 0 & 0
   & 0 & 0 & 0 & 0 \\
 0 & 0 & 0 & 0 & 0 & \tilde g_{14235} & 0 & 0 & 0 & 0 & 0 & 0 & 0 & 0 & 0 & 0 & 0 & 0 & 0 & 0
   & 0 & 0 & 0 & 0 \\
 0 & 0 & 0 & 0 & 0 & 0 & 0 & 0 & 0 & 0 & 0 & 0 & 0 & 0 & 0 & \tilde g_{14325} & 0 & 0 & 0 & 0
   & 0 & 0 & 0 & 0 \\
 0 & 0 & 0 & 0 & 0 & 0 & 0 & 0 & 0 & 0 & 0 & 0 & 0 & 0 & 0 & 0 & 0 & 0 & 0 & 0 & 0 &
   \tilde g_{15432} & 0 & 0 \\
 0 & 0 & 0 & 0 & 0 & 0 & 0 & 0 & 0 & 0 & 0 & 0 & 0 & 0 & 0 & 0 & 0 & 0 & 0 & 0 &
   \tilde g_{14532} & 0 & 0 & 0 \\
 0 & 0 & 0 & 0 & 0 & 0 & 0 & \tilde g_{15342} & 0 & 0 & 0 & 0 & 0 & 0 & 0 & 0 & 0 & 0 & 0 & 0
   & 0 & 0 & 0 & 0 \\
 0 & 0 & 0 & 0 & 0 & 0 & 0 & 0 & 0 & 0 & 0 & 0 & 0 & 0 & 0 & 0 & 0 & 0 & 0 & 0 & 0 & 0 & 0
   & \tilde g_{13542} \\
 0 & 0 & 0 & 0 & 0 & 0 & \tilde g_{14352} & 0 & 0 & 0 & 0 & 0 & 0 & 0 & 0 & 0 & 0 & 0 & 0 & 0
   & 0 & 0 & 0 & 0 \\
 0 & 0 & 0 & 0 & 0 & 0 & 0 & 0 & 0 & 0 & \tilde g_{13452} & 0 & 0 & 0 & 0 & 0 & 0 & 0 & 0 & 0
   & 0 & 0 & 0 & 0 \\
 0 & 0 & 0 & 0 & 0 & 0 & 0 & 0 & 0 & 0 & 0 & 0 & 0 & 0 & 0 & 0 & \tilde g_{15423} & 0 & 0 & 0
   & 0 & 0 & 0 & 0 \\
 0 & 0 & 0 & 0 & 0 & 0 & 0 & 0 & 0 & 0 & 0 & 0 & 0 & 0 & \tilde g_{14523} & 0 & 0 & 0 & 0 & 0
   & 0 & 0 & 0 & 0 \\
 0 & \tilde g_{15243} & 0 & 0 & 0 & 0 & 0 & 0 & 0 & 0 & 0 & 0 & 0 & 0 & 0 & 0 & 0 & 0 & 0 & 0
   & 0 & 0 & 0 & 0 \\
 \tilde g_{14253} & 0 & 0 & 0 & 0 & 0 & 0 & 0 & 0 & 0 & 0 & 0 & 0 & 0 & 0 & 0 & 0 & 0 & 0 & 0
   & 0 & 0 & 0 & 0 \\
 0 & 0 & 0 & 0 & 0 & 0 & 0 & 0 & 0 & 0 & 0 & 0 & 0 & 0 & 0 & 0 & 0 & \tilde g_{15324} & 0 & 0
   & 0 & 0 & 0 & 0 \\
 0 & 0 & 0 & \tilde g_{15234} & 0 & 0 & 0 & 0 & 0 & 0 & 0 & 0 & 0 & 0 & 0 & 0 & 0 & 0 & 0 & 0
   & 0 & 0 & 0 & 0 \\
\end{array}
\right).$}
\end{equation}

\subsection{Solution of the KLT bootstrap}
\label{app:solKLT5pt}

We confirmed explicitly that the solution to the KLT bootstrap at 5-point is reproduced by \autoref{gen5ptleftright},
with 
\begin{align}
    g_{1,\textsc{l}}(1,2,3,4,5) \times m[12345|12345] &= 
    3 \, \frac{e_1}{\Lambda^4}
    - \frac{e_1}{\Lambda^4} \left(\frac{s_{12}}{s_{34}}+\frac{s_{12}}{s_{45}}
    +\text{cyclic}
    \right)
    \nn\\ 
    &  
    +\,\frac{e_2}{\Lambda^6}\,\( 
    -2\,s_{12}+\frac{s_{12}^2}{s_{34}}+\frac{s_{12}^2}{s_{45}}
    +2\,\frac{s_{12}s_{23}}{s_{45}} +\text{cyclic}\)
    \nn\\
     g_{2,\textsc{l}}(1,2,3,4,5) \times m[12345|12345] 
    &= \frac{e_1}{\Lambda^4} 
        - \frac{e_2}{\Lambda^6}\,\left(s_{12}+\text{cyclic}\right)
        \nn\\
    g_{1,\textsc{r}}(1,2,3,4,5) \times m[12345|12345] &= 
    3 \, \frac{e_3}{\Lambda^4}
    - \frac{e_3}{\Lambda^4} \left(\frac{s_{12}}{s_{34}}+\frac{s_{12}}{s_{45}}
    +\text{cyclic}
    \right)
    \nn\\
    g_{2,\textsc{r}}(1,2,3,4,5) \times m[12345|12345] 
    &= \frac{e_3}{\Lambda^4} 
\end{align}
where 
\beq 
m[12345|12345]=\frac{1}{s_{12}s_{34}}+\text{cyclic}\,.
\eeq 
and $e_i$ are free parameters, which reproduce the results of \cite{Chi:2021mio} when 
$e_1=a_{1,0}-a_{1,1}$,
$e_2=a_{2,0}$, and
$e_3=a_{1,1}$.
Note that the $e_2$ parameter could have equivalently been part of $g_{1,\textsc{r}}$ and $g_{2,\textsc{r}}$ without changing the amplitudes.

\subsection{Generalized KLT kernel from seeds}
\label{sec:5pt-kernel-from-seeds}

At 4-point, the generalized kernel for any choice of BCJ bases was obtained by multiplying the left and right side of the traditional KLT kernel by the diagonal matrices ${\bf G}\hd_{\textsc{l}/\textsc{r}}$ for the \emph{same} BCJ bases, see \autoref{examplegenKernel}. 
This simple structure extends to 5-point only for a particular BCJ bases, as the structure gets more involved due the presence of the non-diagonal matrix ${\bf G}_{2}\hd$ in \autoref{gen5ptleftright}. This is not necessarily a problem as one is always allowed to choose a particular basis to compute double-copy amplitudes.

For example, for the usual biadjoint scalar theory, the sub-matrix of ordered amplitudes for 
$\alpha,\beta \in \{12345,13524\}$ reads
\beq 
    m_5[\alpha|\beta] = 
        \pmx{ m_5[12345|12345]& 0\\
             0& m_5[13524|13524]}\,,
\eeq 
where $m_5[12345|13524]=m_5[13524|12345]=0$ because the color orderings $12345$ and $13524$ do not share any poles. The associated kernel is then simply
\beq 
    S_5[\alpha|\beta] = 
        \pmx{ 1/m_5[12345|12345]& 0\\
             0& 1/m_5[13524|13524]}\,.
\eeq 
It turns out that ${\bf m}_5\hd$ is also simplified when restricted to the BCJ basis $\{12345,13524\}$. For any color ordering $a$, and $\delta$ restricted to $\{12345,13524\}$, \autoref{gen5ptleftright} becomes
    \begin{align}\label{mhdindices}
        { m}_5\hd[a|\delta] &= 
        \sum_{b,c}^{24}
        \left( { G}_{1,\textsc{r}} +  G_{2,\textsc{r}}\right)\![a|b] \,
        \; { m}_5[b|c]\,\;
        \left(  G_{1,\textsc{l}} +  G_{2,\textsc{l}}\right)\![c|\delta]
        \,.
    \end{align}
This can be simplified by noting that 
$( { G}_{1,\textsc{l}} +  G_{2,\textsc{l}})[c|12345]$ is nonzero only for
$c=\{12345,14253\}$, while 
$( { G}_{1,\textsc{l}} +  G_{2,\textsc{l}} )[c|13524]$ is nonzero only for
$c=\{12345,13524\}$. Moreover,
$m_5[\alpha|14253]=-m_5[\alpha|13524]$ for any $\alpha$, which means that the second index of $m_5$ can be limited to the BCJ basis  $\{12345,13524\}$. This means that we can restrict to the sub-matrix (appearing in \autoref{eq:curly-g-matrix})
\beq 
~~\mathcal{G}_\textsc{l} 
\equiv \pmx{ \phantom{-}
 ( { G}_{1,\textsc{l}} +  G_{2,\textsc{l}})[12345|12345] &\;
 ( { G}_{1,\textsc{l}} +  G_{2,\textsc{l}})[12345|13524] \\
 -( { G}_{1,\textsc{l}} +  G_{2,\textsc{l}})[14253|12345] &\;
 ( { G}_{1,\textsc{l}} +  G_{2,\textsc{l}})[13524|13524] }\,,
\eeq
and similarly for the right side with
\beq 
~~\mathcal{G}_\textsc{r} 
\equiv \pmx{ 
 ( { G}_{1,\textsc{r}} +  G_{2,\textsc{r}})[12345|12345] &\;
- ( { G}_{1,\textsc{r}} +  G_{2,\textsc{r}})[12345|14253] \\
 ( { G}_{1,\textsc{r}} +  G_{2,\textsc{r}})[14253|12345] &\;\phantom{-}
 ( { G}_{1,\textsc{l}} +  G_{2,\textsc{r}})[13524|13524] }\,,
\eeq
We can then write the BAS amplitudes for this specific BCJ basis as
\beq 
    m_5\hd[\alpha|\delta] = 
    \sum_{\beta,\gamma\in\{12345,13524\}}
        \mathcal{G}_\textsc{r}[\alpha|\beta]
        \cdot     
        m_5[\beta|\gamma]\,
        \cdot
        \mathcal{G}_\textsc{l}[\gamma|\delta]
        \,,
\eeq 
Therefore, in this case, it is simple to invert the generalized BAS matrix to obtain the generalized KLT kernel:
\beq 
    S\hd_5[\alpha|\delta] = 
    \sum_{\beta,\gamma\in\{12345,13524\}} \,
        \mathcal{G}_\textsc{l}^{-1}[\alpha|\beta] \,
        S[\beta|\gamma] \,
        \mathcal{G}_\textsc{r}^{-1}[\gamma|\delta]\,.
\eeq 
We let the investigation of similar formula for a generic BCJ basis for a future work.

\bibliographystyle{JHEP}
\bibliography{refs}

\providecommand{\href}[2]{#2}\begingroup\raggedright\begin{thebibliography}{10}

\bibitem{Kawai:1985xq}
H.~Kawai, D.C.~Lewellen and S.H.H.~Tye, \emph{{A Relation Between Tree
  Amplitudes of Closed and Open Strings}},
  \href{https://doi.org/10.1016/0550-3213(86)90362-7}{\emph{Nucl. Phys. B}
  {\bfseries 269} (1986) 1}.

\bibitem{Bern:2008qj}
Z.~Bern, J.J.M.~Carrasco and H.~Johansson, \emph{{New Relations for
  Gauge-Theory Amplitudes}},
  \href{https://doi.org/10.1103/PhysRevD.78.085011}{\emph{Phys. Rev. D}
  {\bfseries 78} (2008) 085011}
  [\href{https://arxiv.org/abs/0805.3993}{{\ttfamily 0805.3993}}].

\bibitem{Bern:2010ue}
Z.~Bern, J.J.M.~Carrasco and H.~Johansson, \emph{{Perturbative Quantum Gravity
  as a Double Copy of Gauge Theory}},
  \href{https://doi.org/10.1103/PhysRevLett.105.061602}{\emph{Phys. Rev. Lett.}
  {\bfseries 105} (2010) 061602}
  [\href{https://arxiv.org/abs/1004.0476}{{\ttfamily 1004.0476}}].

\bibitem{Cachazo:2013gna}
F.~Cachazo, S.~He and E.Y.~Yuan, \emph{{Scattering equations and
  Kawai-Lewellen-Tye orthogonality}},
  \href{https://doi.org/10.1103/PhysRevD.90.065001}{\emph{Phys. Rev. D}
  {\bfseries 90} (2014) 065001}
  [\href{https://arxiv.org/abs/1306.6575}{{\ttfamily 1306.6575}}].

\bibitem{Cachazo:2013hca}
F.~Cachazo, S.~He and E.Y.~Yuan, \emph{{Scattering of Massless Particles in
  Arbitrary Dimensions}},
  \href{https://doi.org/10.1103/PhysRevLett.113.171601}{\emph{Phys. Rev. Lett.}
  {\bfseries 113} (2014) 171601}
  [\href{https://arxiv.org/abs/1307.2199}{{\ttfamily 1307.2199}}].

\bibitem{Cachazo:2013iea}
F.~Cachazo, S.~He and E.Y.~Yuan, \emph{{Scattering of Massless Particles:
  Scalars, Gluons and Gravitons}},
  \href{https://doi.org/10.1007/JHEP07(2014)033}{\emph{JHEP} {\bfseries 07}
  (2014) 033} [\href{https://arxiv.org/abs/1309.0885}{{\ttfamily 1309.0885}}].

\bibitem{Carrasco:2019qwr}
J.J.M.~Carrasco and L.~Rodina, \emph{{UV considerations on scattering
  amplitudes in a web of theories}},
  \href{https://doi.org/10.1103/PhysRevD.100.125007}{\emph{Phys. Rev. D}
  {\bfseries 100} (2019) 125007}
  [\href{https://arxiv.org/abs/1908.08033}{{\ttfamily 1908.08033}}].

\bibitem{Johansson:2014zca}
H.~Johansson and A.~Ochirov, \emph{{Pure Gravities via Color-Kinematics Duality
  for Fundamental Matter}},
  \href{https://doi.org/10.1007/JHEP11(2015)046}{\emph{JHEP} {\bfseries 11}
  (2015) 046} [\href{https://arxiv.org/abs/1407.4772}{{\ttfamily 1407.4772}}].

\bibitem{Johansson:2015oia}
H.~Johansson and A.~Ochirov, \emph{{Color-Kinematics Duality for QCD
  Amplitudes}}, \href{https://doi.org/10.1007/JHEP01(2016)170}{\emph{JHEP}
  {\bfseries 01} (2016) 170}
  [\href{https://arxiv.org/abs/1507.00332}{{\ttfamily 1507.00332}}].

\bibitem{delaCruz:2016wbr}
L.~de~la Cruz, A.~Kniss and S.~Weinzierl, \emph{{Double Copies of Fermions as
  Matter that Interacts Only Gravitationally}},
  \href{https://doi.org/10.1103/PhysRevLett.116.201601}{\emph{Phys. Rev. Lett.}
  {\bfseries 116} (2016) 201601}
  [\href{https://arxiv.org/abs/1601.04523}{{\ttfamily 1601.04523}}].

\bibitem{Brown:2016hck}
R.W.~Brown and S.G.~Naculich, \emph{{Color-factor symmetry and BCJ relations
  for QCD amplitudes}},
  \href{https://doi.org/10.1007/JHEP11(2016)060}{\emph{JHEP} {\bfseries 11}
  (2016) 060} [\href{https://arxiv.org/abs/1608.05291}{{\ttfamily
  1608.05291}}].

\bibitem{Brown:2018wss}
R.W.~Brown and S.G.~Naculich, \emph{{KLT-type relations for QCD and bicolor
  amplitudes from color-factor symmetry}},
  \href{https://doi.org/10.1007/JHEP03(2018)057}{\emph{JHEP} {\bfseries 03}
  (2018) 057} [\href{https://arxiv.org/abs/1802.01620}{{\ttfamily
  1802.01620}}].

\bibitem{Johansson:2019dnu}
H.~Johansson and A.~Ochirov, \emph{{Double copy for massive quantum particles
  with spin}}, \href{https://doi.org/10.1007/JHEP09(2019)040}{\emph{JHEP}
  {\bfseries 09} (2019) 040}
  [\href{https://arxiv.org/abs/1906.12292}{{\ttfamily 1906.12292}}].

\bibitem{Plefka:2019wyg}
J.~Plefka, C.~Shi and T.~Wang, \emph{{Double copy of massive scalar QCD}},
  \href{https://doi.org/10.1103/PhysRevD.101.066004}{\emph{Phys. Rev. D}
  {\bfseries 101} (2020) 066004}
  [\href{https://arxiv.org/abs/1911.06785}{{\ttfamily 1911.06785}}].

\bibitem{Chiodaroli:2015rdg}
M.~Chiodaroli, M.~Gunaydin, H.~Johansson and R.~Roiban, \emph{{Spontaneously
  Broken Yang-Mills-Einstein Supergravities as Double Copies}},
  \href{https://doi.org/10.1007/JHEP06(2017)064}{\emph{JHEP} {\bfseries 06}
  (2017) 064} [\href{https://arxiv.org/abs/1511.01740}{{\ttfamily
  1511.01740}}].

\bibitem{Johnson:2020pny}
L.A.~Johnson, C.R.T.~Jones and S.~Paranjape, \emph{{Constraints on a Massive
  Double-Copy and Applications to Massive Gravity}},
  \href{https://doi.org/10.1007/JHEP02(2021)148}{\emph{JHEP} {\bfseries 02}
  (2021) 148} [\href{https://arxiv.org/abs/2004.12948}{{\ttfamily
  2004.12948}}].

\bibitem{Momeni:2020vvr}
A.~Momeni, J.~Rumbutis and A.J.~Tolley, \emph{{Massive Gravity from Double
  Copy}}, \href{https://doi.org/10.1007/JHEP12(2020)030}{\emph{JHEP} {\bfseries
  12} (2020) 030} [\href{https://arxiv.org/abs/2004.07853}{{\ttfamily
  2004.07853}}].

\bibitem{Momeni:2020hmc}
A.~Momeni, J.~Rumbutis and A.J.~Tolley, \emph{{Kaluza-Klein from
  colour-kinematics duality for massive fields}},
  \href{https://doi.org/10.1007/JHEP08(2021)081}{\emph{JHEP} {\bfseries 08}
  (2021) 081} [\href{https://arxiv.org/abs/2012.09711}{{\ttfamily
  2012.09711}}].

\bibitem{Hang:2021fmp}
Y.-F.~Hang and H.-J.~He, \emph{{Structure of Kaluza-Klein graviton scattering
  amplitudes from the gravitational equivalence theorem and double copy}},
  \href{https://doi.org/10.1103/PhysRevD.105.084005}{\emph{Phys. Rev. D}
  {\bfseries 105} (2022) 084005}
  [\href{https://arxiv.org/abs/2106.04568}{{\ttfamily 2106.04568}}].

\bibitem{Li:2021yfk}
Y.~Li, Y.-F.~Hang, H.-J.~He and S.~He, \emph{{Scattering amplitudes of
  Kaluza-Klein strings and extended massive double-copy}},
  \href{https://doi.org/10.1007/JHEP02(2022)120}{\emph{JHEP} {\bfseries 02}
  (2022) 120} [\href{https://arxiv.org/abs/2111.12042}{{\ttfamily
  2111.12042}}].

\bibitem{Chen:2013fya}
G.~Chen and Y.-J.~Du, \emph{{Amplitude Relations in Non-linear Sigma Model}},
  \href{https://doi.org/10.1007/JHEP01(2014)061}{\emph{JHEP} {\bfseries 01}
  (2014) 061} [\href{https://arxiv.org/abs/1311.1133}{{\ttfamily 1311.1133}}].

\bibitem{Cachazo:2014xea}
F.~Cachazo, S.~He and E.Y.~Yuan, \emph{{Scattering Equations and Matrices: From
  Einstein To Yang-Mills, DBI and NLSM}},
  \href{https://doi.org/10.1007/JHEP07(2015)149}{\emph{JHEP} {\bfseries 07}
  (2015) 149} [\href{https://arxiv.org/abs/1412.3479}{{\ttfamily 1412.3479}}].

\bibitem{Bern:2019prr}
Z.~Bern, J.J.~Carrasco, M.~Chiodaroli, H.~Johansson and R.~Roiban, \emph{{The
  Duality Between Color and Kinematics and its Applications}},
  \href{https://arxiv.org/abs/1909.01358}{{\ttfamily 1909.01358}}.

\bibitem{Broedel:2012rc}
J.~Broedel and L.J.~Dixon, \emph{{Color-kinematics duality and double-copy
  construction for amplitudes from higher-dimension operators}},
  \href{https://doi.org/10.1007/JHEP10(2012)091}{\emph{JHEP} {\bfseries 10}
  (2012) 091} [\href{https://arxiv.org/abs/1208.0876}{{\ttfamily 1208.0876}}].

\bibitem{Menezes:2021dyp}
G.~Menezes, \emph{{Color-kinematics duality, double copy and the unitarity
  method for higher-derivative QCD and quadratic gravity}},
  \href{https://doi.org/10.1007/JHEP03(2022)074}{\emph{JHEP} {\bfseries 03}
  (2022) 074} [\href{https://arxiv.org/abs/2112.00978}{{\ttfamily
  2112.00978}}].

\bibitem{Elvang:2018dco}
H.~Elvang, M.~Hadjiantonis, C.R.T.~Jones and S.~Paranjape, \emph{{Soft
  Bootstrap and Supersymmetry}},
  \href{https://doi.org/10.1007/JHEP01(2019)195}{\emph{JHEP} {\bfseries 01}
  (2019) 195} [\href{https://arxiv.org/abs/1806.06079}{{\ttfamily
  1806.06079}}].

\bibitem{Elvang:2020kuj}
H.~Elvang, M.~Hadjiantonis, C.R.T.~Jones and S.~Paranjape,
  \emph{{Electromagnetic Duality and D3-Brane Scattering Amplitudes Beyond
  Leading Order}}, \href{https://doi.org/10.1007/JHEP04(2021)173}{\emph{JHEP}
  {\bfseries 04} (2021) 173}
  [\href{https://arxiv.org/abs/2006.08928}{{\ttfamily 2006.08928}}].

\bibitem{CarrilloGonzalez:2019fzc}
M.~Carrillo~Gonz\'alez, R.~Penco and M.~Trodden, \emph{{Shift symmetries, soft
  limits, and the double copy beyond leading order}},
  \href{https://doi.org/10.1103/PhysRevD.102.105011}{\emph{Phys. Rev. D}
  {\bfseries 102} (2020) 105011}
  [\href{https://arxiv.org/abs/1908.07531}{{\ttfamily 1908.07531}}].

\bibitem{Kampf:2021jvf}
K.~Kampf, \emph{{The ChPT: top-down and bottom-up}},
  \href{https://doi.org/10.1007/JHEP12(2021)140}{\emph{JHEP} {\bfseries 12}
  (2021) 140} [\href{https://arxiv.org/abs/2109.11574}{{\ttfamily
  2109.11574}}].

\bibitem{Low:2019wuv}
I.~Low and Z.~Yin, \emph{{New Flavor-Kinematics Dualities and Extensions of
  Nonlinear Sigma Models}},
  \href{https://doi.org/10.1016/j.physletb.2020.135544}{\emph{Phys. Lett. B}
  {\bfseries 807} (2020) 135544}
  [\href{https://arxiv.org/abs/1911.08490}{{\ttfamily 1911.08490}}].

\bibitem{Low:2020ubn}
I.~Low, L.~Rodina and Z.~Yin, \emph{{Double Copy in Higher Derivative Operators
  of Nambu-Goldstone Bosons}},
  \href{https://doi.org/10.1103/PhysRevD.103.025004}{\emph{Phys. Rev. D}
  {\bfseries 103} (2021) 025004}
  [\href{https://arxiv.org/abs/2009.00008}{{\ttfamily 2009.00008}}].

\bibitem{Brandhuber:2021kpo}
A.~Brandhuber, G.~Chen, G.~Travaglini and C.~Wen, \emph{{A new gauge-invariant
  double copy for heavy-mass effective theory}},
  \href{https://doi.org/10.1007/JHEP07(2021)047}{\emph{JHEP} {\bfseries 07}
  (2021) 047} [\href{https://arxiv.org/abs/2104.11206}{{\ttfamily
  2104.11206}}].

\bibitem{Haddad:2020tvs}
K.~Haddad and A.~Helset, \emph{{The double copy for heavy particles}},
  \href{https://doi.org/10.1103/PhysRevLett.125.181603}{\emph{Phys. Rev. Lett.}
  {\bfseries 125} (2020) 181603}
  [\href{https://arxiv.org/abs/2005.13897}{{\ttfamily 2005.13897}}].

\bibitem{Carrasco:2019yyn}
J.J.M.~Carrasco, L.~Rodina, Z.~Yin and S.~Zekioglu, \emph{{Simple encoding of
  higher derivative gauge and gravity counterterms}},
  \href{https://doi.org/10.1103/PhysRevLett.125.251602}{\emph{Phys. Rev. Lett.}
  {\bfseries 125} (2020) 251602}
  [\href{https://arxiv.org/abs/1910.12850}{{\ttfamily 1910.12850}}].

\bibitem{Carrasco:2021ptp}
J.J.M.~Carrasco, L.~Rodina and S.~Zekioglu, \emph{{Composing effective
  prediction at five points}},
  \href{https://doi.org/10.1007/JHEP06(2021)169}{\emph{JHEP} {\bfseries 06}
  (2021) 169} [\href{https://arxiv.org/abs/2104.08370}{{\ttfamily
  2104.08370}}].

\bibitem{Chi:2021mio}
H.-H.~Chi, H.~Elvang, A.~Herderschee, C.R.T.~Jones and S.~Paranjape,
  \emph{{Generalizations of the double-copy: the KLT bootstrap}},
  \href{https://doi.org/10.1007/JHEP03(2022)077}{\emph{JHEP} {\bfseries 03}
  (2022) 077} [\href{https://arxiv.org/abs/2106.12600}{{\ttfamily
  2106.12600}}].

\bibitem{Mizera:2016jhj}
S.~Mizera, \emph{{Inverse of the String Theory KLT Kernel}},
  \href{https://doi.org/10.1007/JHEP06(2017)084}{\emph{JHEP} {\bfseries 06}
  (2017) 084} [\href{https://arxiv.org/abs/1610.04230}{{\ttfamily
  1610.04230}}].

\bibitem{Bjerrum-Bohr:2010pnr}
N.E.J.~Bjerrum-Bohr, P.H.~Damgaard, T.~Sondergaard and P.~Vanhove, \emph{{The
  Momentum Kernel of Gauge and Gravity Theories}},
  \href{https://doi.org/10.1007/JHEP01(2011)001}{\emph{JHEP} {\bfseries 01}
  (2011) 001} [\href{https://arxiv.org/abs/1010.3933}{{\ttfamily 1010.3933}}].

\bibitem{KiermaierTalk2010}
M.~Kiermaier, \emph{Gravity as the square of gauge theory}, {\emph{{\rm
  presentation at}
  \href{https://strings.ph.qmul.ac.uk/~theory/Amplitudes2010/Talks/MK2010.pdf}{``Amplitudes
  2010''}} (7 May 2010) \!\!.}

\bibitem{Bern:2011ia}
Z.~Bern and T.~Dennen, \emph{{A Color Dual Form for Gauge-Theory Amplitudes}},
  \href{https://doi.org/10.1103/PhysRevLett.107.081601}{\emph{Phys. Rev. Lett.}
  {\bfseries 107} (2011) 081601}
  [\href{https://arxiv.org/abs/1103.0312}{{\ttfamily 1103.0312}}].

\bibitem{Bjerrum-Bohr:2012kaa}
N.E.J.~Bjerrum-Bohr, P.H.~Damgaard, R.~Monteiro and D.~O'Connell,
  \emph{{Algebras for Amplitudes}},
  \href{https://doi.org/10.1007/JHEP06(2012)061}{\emph{JHEP} {\bfseries 06}
  (2012) 061} [\href{https://arxiv.org/abs/1203.0944}{{\ttfamily 1203.0944}}].

\bibitem{Du:2013sha}
Y.-J.~Du, B.~Feng and C.-H.~Fu, \emph{{The Construction of Dual-trace Factor in
  Yang-Mills Theory}},
  \href{https://doi.org/10.1007/JHEP07(2013)057}{\emph{JHEP} {\bfseries 07}
  (2013) 057} [\href{https://arxiv.org/abs/1304.2978}{{\ttfamily 1304.2978}}].

\bibitem{Fu:2013qna}
C.-H.~Fu, Y.-J.~Du and B.~Feng, \emph{{Note on Construction of Dual-trace
  Factor in Yang-Mills Theory}},
  \href{https://doi.org/10.1007/JHEP10(2013)069}{\emph{JHEP} {\bfseries 10}
  (2013) 069} [\href{https://arxiv.org/abs/1305.2996}{{\ttfamily 1305.2996}}].

\bibitem{Naculich:2014rta}
S.G.~Naculich, \emph{{Scattering equations and virtuous kinematic numerators
  and dual-trace functions}},
  \href{https://doi.org/10.1007/JHEP07(2014)143}{\emph{JHEP} {\bfseries 07}
  (2014) 143} [\href{https://arxiv.org/abs/1404.7141}{{\ttfamily 1404.7141}}].

\bibitem{Brandhuber:2021bsf}
A.~Brandhuber, G.~Chen, H.~Johansson, G.~Travaglini and C.~Wen,
  \emph{{Kinematic Hopf Algebra for Bern-Carrasco-Johansson Numerators in
  Heavy-Mass Effective Field Theory and Yang-Mills Theory}},
  \href{https://doi.org/10.1103/PhysRevLett.128.121601}{\emph{Phys. Rev. Lett.}
  {\bfseries 128} (2022) 121601}
  [\href{https://arxiv.org/abs/2111.15649}{{\ttfamily 2111.15649}}].

\bibitem{KLEISS1989616}
R.~Kleiss and H.~Kuijf, \emph{Multigluon cross sections and 5-jet production at
  hadron colliders},
  \href{https://doi.org/https://doi.org/10.1016/0550-3213(89)90574-9}{\emph{Nuclear
  Physics B} {\bfseries 312} (1989) 616}.

\bibitem{elvangQCDmeetsGravity2021}
H.~Elvang, \emph{Effective field theories and the double-copy}, {\emph{{\rm
  presentation at}
  \href{https://bhaumik-institute.physics.ucla.edu/QCD2021}{``QCD meets Gravity
  2021''}} (15 December 2021) \!\!.}

\bibitem{Elvang:2013cua}
H.~Elvang and Y.-t.~Huang, \emph{{Scattering Amplitudes}},
  \href{https://arxiv.org/abs/1308.1697}{{\ttfamily 1308.1697}}.

\bibitem{Cheung:2017pzi}
C.~Cheung, \emph{{TASI Lectures on Scattering Amplitudes}},  in
  \emph{{Proceedings, Theoretical Advanced Study Institute in Elementary
  Particle Physics : Anticipating the Next Discoveries in Particle Physics
  (TASI 2016)}: {Boulder, CO, USA, June 6-July 1, 2016}}, R.~Essig and I.~Low,
  eds., pp.~571--623 (2018),
  \href{https://doi.org/10.1142/9789813233348_0008}{DOI}
  [\href{https://arxiv.org/abs/1708.03872}{{\ttfamily 1708.03872}}].

\bibitem{DelDuca:1999rs}
V.~Del~Duca, L.J.~Dixon and F.~Maltoni, \emph{{New color decompositions for
  gauge amplitudes at tree and loop level}},
  \href{https://doi.org/10.1016/S0550-3213(99)00809-3}{\emph{Nucl. Phys. B}
  {\bfseries 571} (2000) 51}
  [\href{https://arxiv.org/abs/hep-ph/9910563}{{\ttfamily hep-ph/9910563}}].

\bibitem{deNeeling:2022tsu}
D.~de~Neeling, D.~Roest and S.~Veldmeijer, \emph{{Flavour-kinematic duality for
  Goldstone modes}},  \href{https://arxiv.org/abs/2204.11629}{{\ttfamily
  2204.11629}}.

\bibitem{Boels:2016xhc}
R.H.~Boels and R.~Medina, \emph{{Graviton and gluon scattering from first
  principles}},
  \href{https://doi.org/10.1103/PhysRevLett.118.061602}{\emph{Phys. Rev. Lett.}
  {\bfseries 118} (2017) 061602}
  [\href{https://arxiv.org/abs/1607.08246}{{\ttfamily 1607.08246}}].

\bibitem{Bern:2017tuc}
Z.~Bern, A.~Edison, D.~Kosower and J.~Parra-Martinez, \emph{{Curvature-squared
  multiplets, evanescent effects, and the U(1) anomaly in $N=4$ supergravity}},
  \href{https://doi.org/10.1103/PhysRevD.96.066004}{\emph{Phys. Rev. D}
  {\bfseries 96} (2017) 066004}
  [\href{https://arxiv.org/abs/1706.01486}{{\ttfamily 1706.01486}}].

\bibitem{Ben-Israel1980}
A.~Ben-Israel and T.N.~Greville, \emph{{Generalized Inverses: Theory and
  Applications}}, Wiley, New York (1980).

\end{thebibliography}\endgroup
\end{document}